\pdfoutput=1
\documentclass[useAMS,usenatbib,a4paper]{mn2e}
\usepackage{times}
\usepackage{ifthen}
\usepackage{paralist}
\usepackage{xfrac}
\newboolean{draft}
\newboolean{hyperlinks}
\newboolean{thesis}
\setboolean{draft}{false}
\setboolean{hyperlinks}{true}
\setboolean{thesis}{false} % boolvar=true or false

\usepackage{amsmath}
\usepackage{natbib} %bibtex
\usepackage{color} %color text
\usepackage{array}
\usepackage{booktabs}
\usepackage{graphicx}
\usepackage{multirow} % multiple rows in tables
\usepackage{rotating} % to rotate figures
\usepackage{framed} %to make frames
\usepackage{tabularx}% for adding a little bit of extra vertical space to tables
\usepackage{longtable}
\usepackage{booktabs}
\usepackage{etoolbox} % for boolean toggles
\usepackage{enumitem}% for more control of enumerate
\usepackage{ctable}
\usepackage{subfigure}
\usepackage{rotating}
\usepackage{xspace}

\usepackage{ulem} % for strikeout text

\ifthenelse{\boolean{hyperlinks}}
{
  \usepackage{hyperref} % to set up hyperrefs within paper
  \usepackage[all]{hypcap} %keeps hyperrefs to figures leading to the whole plot instead of just caption.
}{}

\ifthenelse {\boolean{thesis}}
{
\usepackage{nonfloat}
}
{
} % list of added packages

\let\jnlstyle=\rm
\def\refjnl#1{{\jnlstyle#1}}

\def\aj{\refjnl{AJ}}                   % Astronomical Journal
\def\araa{\refjnl{ARA\&A}}             % Annual Review of Astron and Astrophys
\def\apj{\refjnl{ApJ}}                 % Astrophysical Journal
\def\apjl{\refjnl{ApJL}}                % Astrophysical Journal, Letters
\def\apjs{\refjnl{ApJS}}               % Astrophysical Journal, Supplement
\def\aap{\refjnl{A\&A}}                % Astronomy and Astrophysics
\def\mnras{\refjnl{MNRAS}}             % Monthly Notices of the RAS
\def\nar{\refjnl{New A Rev.}}          % New Astronomy Review
\def\prd{\refjnl{Phys.~Rev.~D}}        % Physical Review D
\def\nat{\refjnl{Nature}}              % Nature
\def\planss{\refjnl{Planet.~Space~Sci.}}   % Planetary Space Science
 \newboolean{outline}
 \newboolean{notes}
 \setboolean{outline}{false}
 \setboolean{notes}{true}
 \def\2F1{\mbox{$_2${F}$_1$}}
 
 \newcommand{\arr}{\ensuremath{a_{\rm rr}}\xspace}
 \newcommand{\asec}{\ensuremath{a_\bullet}\xspace}
 \newcommand{\astar}{\ensuremath{a_\star}\xspace}
 
 \def\BA#1\EA{\begin{align}#1\end{align}}
 \newcommand{\BE}{\begin{equation} }
 \newcommand{\BEn}{\begin{enumerate}[leftmargin=0.5cm]}
 \newcommand{\BI}{\begin{itemize}}
 \newcommand{\BEA}{\begin{eqnarray}}

 \newcommand{\roughdraft}[2]{\ifthenelse{\boolean{draft}}{#1}{#2}}
 \newcommand{\EE}{\end{equation} } 
 \newcommand{\EEA}{\end{eqnarray}}
 \newcommand{\EEn}{\end{enumerate}}
 \newcommand{\EI}{\end{itemize}}

 \newcommand{\eqn}[1]{Eqn.~\ref{#1}}
 \newcommand{\Eqn}[1]{Eqn.~\ref{#1}} 
 \newcommand{\eqns}[2]{Eqns.~\ref{#1}--\ref{#2}}
 \newcommand{\eqnsand}[2]{Eqns.~\ref{#1} and \ref{#2}}

 \newcommand{\fig}[1]{Fig.~\ref{#1}} 
 \newcommand{\fignum}[1]{Fig.~{#1}}
 \newcommand{\figlistfive}[5]{Figs.~\ref{#1}, \ref{#2}, \ref{#3}, \ref{#4}, and \ref{#5}}
 \newcommand{\figlistfour}[4]{Figs.~\ref{#1}, \ref{#2}, \ref{#3}, and \ref{#4}}
 \newcommand{\figlistthree}[3]{Figs.~\ref{#1}, \ref{#2}, and \ref{#3}}
 
 \newcommand{\figsand}[2]{Figs.~\ref{#1} and \ref{#2}}
 \newcommand{\Gpc}{\,\text{Gpc}}
 
 \newcommand{\gsim}{\mathrel{\hbox{\rlap{\lower.55ex\hbox{$\sim$}} \kern-.3em \raise.4ex \hbox{$>$}}}}

 \newcommand{\km}{{\rm km}}
 
 \newcommand{\Lc}{\ensuremath{L_{\rm c}}\xspace}
 
 \newcommand{\Lbin}{\ensuremath{L}_{\rm b}\xspace}
 \newcommand{\Lplunge}{\ensuremath{L_{\rm plunge}}\xspace}
 \newcommand{\lsim}{\mathrel{\hbox{\rlap{\lower.55ex\hbox{$\sim$}} \kern-.3em \raise.4ex \hbox{$<$}}}}
 \newcommand{\Lstar}{\ensuremath{L_\star}\xspace}
 \newcommand{\Ltot}{\ensuremath{L}\xspace}
 \newcommand{\Lvec}{\ensuremath{\vec{L}}\xspace}
 \newcommand{\Lz}{\ensuremath{L_{\rm z}}\xspace}

 \newcommand{\mprim}{\ensuremath{M}\xspace}
 \newcommand{\msec}{\ensuremath{M_\bullet}\xspace}
 
 \newcommand{\Mstar}{M_\star}
 \newcommand{\mstar}{m_\star}
 \newcommand{\msun}{\ensuremath{M_{\odot}}\xspace}
 \newcommand{\Myr}{\;\ensuremath{{\rm Myr}}\xspace}
 
 \newcommand{\nframe}[1]{\ifthenelse{\boolean{outline}}{{{\color{red}\begin{framed}{\begin{enumerate} #1\end{enumerate}}\end{framed}}}}}{}
 \newcommand{\nn}{\nonumber}
 \newcommand{\note}[1]{\ifthenelse{\boolean{notes}}{\textcolor{red}{(#1)}}\ }{}
 
 \providecommand{\OO}[1]{{\mathcal O} \left(#1\right)}

 \newcommand{\pc}{{\rm\, pc}}

 \newcommand{\psec}{\ensuremath{P_\bullet}\xspace}
 \newcommand{\pstar}{\ensuremath{P_\star}\xspace}
 \def\r{\mathbf{r}}

 \newcommand{\rhostar}{\rho_\star}

 \newcommand{\rs}{r_{\rm S}}
 
 \newcommand{\rstall}{\ensuremath{R_{\rm Stall}\xspace}}

 \newcommand{\s}{\,\rm s}
 
 \newcommand{\sect}[1]{Sec.~\ref{#1}}
 \newcommand{\sects}[2]{{Secs.~\ref{#1}--\ref{#2}}}
 \newcommand{\sectsand}[2]{{Secs.~\ref{#1} and \ref{#2}}}

 \newcommand{\tab}[1]{Table~\ref{#1}}

 \newcommand{\texterior}{\ensuremath{t_{\rm\phi,ext}}\xspace}

 \newcommand{\tkoz}{\ensuremath{t_{\rm Kozai}}\xspace}
 \newcommand{\tKoz}{\ensuremath{T_\text{Kozai}}\xspace}
 \newcommand{\tL}{\tilde{L}}

 \newcommand{\trp}{\tilde{r}_{\rm p}}

 \newcommand{\tphigr}{\ensuremath{t_{\rm\phi,GR}}\xspace}
 \newcommand{\tphisp}{\ensuremath{t_{\rm\phi,SP}}\xspace}
 \newcommand{\trr}{\ensuremath{t_{\rm rr}}\xspace}

 \def\v{\mathbf{v}}

 \newcommand{\yr}{{\rm yr}}

\ifthenelse{\boolean{thesis}}{
   \newcommand{\chap}{\chapter}

}{
   \newcommand{\chap}{\title}
   \newcommand{\chapter}{\title}

}

\providecommand{\supportfrom}[1]{#1}

\setboolean{outline}{false}
\newtoggle{check}
\newtoggle{comments}
\toggletrue{check}
\toggletrue{comments}

\newcommand{\tdpaper}{{WB11}\xspace}

\chap[EMRIs due to SMBH Binaries]{Production of EMRIs in Supermassive Black Hole Binaries} 
\label{sec:5:title}
 
\author[Bode \& Wegg]
{J.~Nate\ Bode$^{1,2}$\thanks{E-mail: natebode@gmail.com} and Christopher Wegg$^{1,3}$\thanks{E-mail: wegg@mpe.mpg.de}\\
$^1${Theoretical Astrophysics, California Institute of Technology, M/C 350-17, Pasadena, CA 91125.}\\
$^2${Boston Consulting Group, 300 N. LaSalle Ave, Chicago, Ill, 60654}\\
$^3${Max-Planck-Institut f\"ur Extraterrestrische Physik, Giessenbachstrasse, 85748 Garching, Germany.}}

\pagerange{\pageref{firstpage}--\pageref{lastpage}} \pubyear{2012}
 
\begin{document}

\label{firstpage}

\maketitle

%-----------------------------------ABSTRACT--------------------------------------
\begin{abstract}
We consider the formation of extreme mass-ratio inspirals (EMRIs) sourced from a stellar cusp centred on a primary supermassive black hole (SMBH) and perturbed by an inspiraling less massive secondary SMBH. The problem is approached numerically, assuming the stars are non-interacting over these short timescales and performing an ensemble of restricted three-body integrations. From these simulations we see that not only can EMRIs be produced during this process, but the dynamics are also quite rich. In particular, most of the EMRIs are produced through a process akin to the Kozai-Lidov mechanism, but with strong effects due to the non-Keplerian stellar potential, general relativity, and non-secular oscillations in the angular momentum on the orbital timescale of the binary SMBH system.

\end{abstract}

%---------------------------------------------------------------------------------

%------------------------------------BODY-----------------------------------------

%=============================================
\section{Introduction}
\label{sec:5:introduction}
%=============================================
One of the most interesting sources for low frequency gravitational wave (GW) detectors such as the final incarnation of the {\it Laser Interferometer Space Antenna} (LISA), or its newer scion European LISA (eLISA), is the capture of stellar mass compact objects (COs) by a supermassive black hole (SMBH). COs are the final state of stellar evolution and include stellar mass black holes, neutron stars, and white dwarfs. Due to the significant mass difference between the SMBH and the inspiraling CO, these sources are referred to as extreme mass-ratio inspirals (EMRIs).

Detection of EMRIs by GW detectors provides: 1) an accurate measurement of the spin and mass of the SMBH \citep{Barack:Cutler:04} along with a moderate determination of its location, 2) a test that the spin and mass are the only parameters characterizing the black hole's space-time \citep[termed `bothrodesy';][]{Ryan:97,Hughes:09}, 3) information about the presence of a secondary SMBH orbiting the primary SMBH \citep{Yunes+10},  4) information about the presence of a gaseous disk in the system \citep{Narayan:00,Yunes+11}, and 5) a possible electromagnetic counterpart to the LISA signal \citep{Sesana+08,Menou+08} if the source was a white dwarf. Such an electromagnetic counterpart would localise the host, thus allowing for follow up observation.

\nframe{
\item EMRIs and IMRIs
\BEn 
\item characteristic distances and sizes
\item we consider EMRIs
\item IMRIs undergo similar process, less numerous
\EEn
}

The production of EMRIs amounts to either forming COs on, or driving them onto, orbits whose GW inspiral time is shorter than the timescale for other orbital perturbations. The standard method of EMRI production \citep{Hils:Bender:1995} is that COs are transported to the EMRI loss cone via gravitational scattering with other stellar mass objects. This is an improbable event because, as the CO's orbit becomes more eccentric and the rate of orbital energy loss to GW emission increases, ever smaller kicks to its angular momentum may remove it from this orbit or plunge it directly into the central SMBH. Despite the apparent unlikeliness of this process, many such EMRIs are expected to form, and be observable by low frequency gravitational wave missions \citep[see][for a review of the subject]{AmaroSeoane+07}. 

Other possible EMRI formation mechanisms include the in situ formation of COs via a massive self-gravitating accretion disk \citep{Levin:03} like that which is believed to have existed in the Milky Way \citep{Levin:Beloborodov:03}. Alternatively, the CO can be deposited close to the SMBH by a stellar binary which interacts strongly with the SMBH and ejects the CO's partner, while leaving the CO on a low eccentricity orbit with small semi-major axis \citep{Miller+05}. 

We consider a different scattering method; one where a secondary SMBH is present and entering the final stage of its inspiral \citep{Begelman:80,Milosavljevic:Merritt:03} due to dynamical friction. In this scenario the scattering phase is short-lived, but the rate of stars scattered to highly eccentric orbits is significantly increased. Moreover, the secondary SMBH induces Lidov-Kozai oscillations in the orbital elements of many stars, considered first in the context of tidal disruptions by \citet{Ivanov+05}. Because some of these stars are COs, on passages close to the primary SMBH, they are not tidally disrupted, but instead radiate a fraction of their orbital energy in gravitational waves, leaving for the possibility of ultimately becoming an EMRI. 

We investigate this possibility by performing an ensemble of three body integrations. The equations of motion of a star randomly chosen from a stellar cusp centred on a primary supermassive black hole (SMBH) are integrated in the presence of a secondary SMBH inspiraling on a pre-calculated path determined by dynamical friction (\sect{sec:5:secondaryinspiral}). We consider a $10^6\msun$ primary SMBH and various masses of the secondary SMBH and stars. The stars are then followed to determine if they eventually become EMRIs.  For clarity we add that throughout the epochs of SMBH binary evolution we consider GW radiation has no practical effect on the SMBH orbit.

We present the results as follows. In \sect{sec:5:background} we present a brief introduction to EMRI formation under the standard channel, while we describe the assumptions and physical setup in \sect{sec:5:physicalsetup}. The simulation is described in \sect{sec:5:simulation} and the resulting rates are provided in \sect{sec:5:emrirates}. To understand the dynamics which result in the majority of our EMRIs we elucidate the standard Kozai-Lidov mechanism, as well as extensions, in \sect{sec:5:oom}. Our rates and the relevance of the assumptions they are based on are discussed in \sect{sec:5:discussion}. In \sect{sec:5:conclusion} we present our conclusions.

%=============================================
\section{Background}
\label{sec:5:background}
%=============================================

%--------------------------------
\begin{figure*}
\begin{center}
\includegraphics[width=\textwidth]{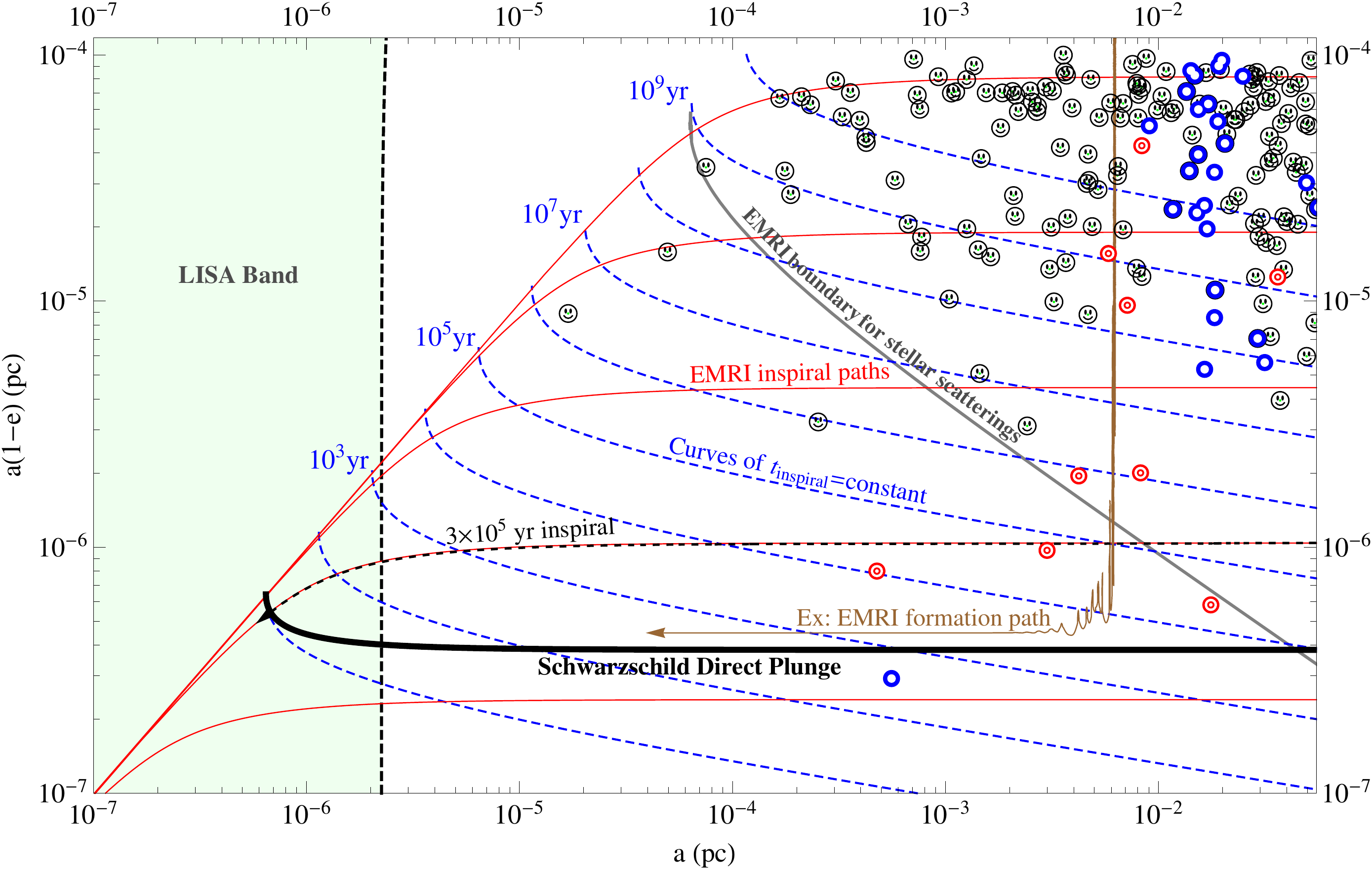}
\caption[The EMRI parameter space with new addition]{\label{fig:5:emriparameterspace}
The standard EMRI parameter space with new addition: We plot  $a(1-e)$ as a function of $a$ with $a$ and $e$ defined by \eqnsand{eqn:5:aofE}{eqn:5:eofLa}, respectively. A compact object (stellar-mass black hole, neutron star, white dwarf) inspirals due to GW radiation along paths shown in solid red and approximated using the far-field equations of \cite{Peters:64}. The timescale for inspiral is approximately given by the times on the dashed blue curves, which are also calculated from \cite{Peters:64}. However, if no secondary SMBH is present and if the initial $\{a$, $a(1-e)\}$ pair lies above the solid gray line, the star is unlikely to complete its inspiral before two-body stellar scatterings move the orbit to larger $a(1-e)$. The minimum angular momentum a star can have while on a parabolic orbit without plunging into the primary SMBH, \Lplunge (also referred to as the unstable circular orbit or separatrix), is plotted in thick solid black. In our simulation with mass ratio $q=0.3$ and stellar masses $\mstar=10\msun$ the red targets are the initial conditions of a star which turned into an EMRI, the thick blue circles are the initial conditions for stars which turn into plunges, and the smiley faces are the initial conditions for stars who end the simulation uneventfully. We use the same initial conditions for all of our simulations, though which stars become EMRIs, plunges, or neither obviously differs.
As an example of a  star formed from the presence of a secondary SMBH we show in solid brown the path of one of the EMRIs produced in our simulations. The star is also marked as a red-outlined gold star in \figlistthree{fig:5:kozaijc}{fig:5:kozai}{fig:5:oom} and used to demonstrate the importance of precession due to the stellar potential and GR in \fig{fig:5:revkozaiprecession}. This star has a path similar to many seen in our simulations, elucidating the new channel of EMRI production.
}
\end{center}
\end{figure*}
%--------------------------------

%----------------------------------------------------------------------------
\subsection{Parameter Space}
%----------------------------------------------------------------------------

Many of the key aspects of the standard method of EMRI formation are illustrated in \fig{fig:5:emriparameterspace} where we plot $a(1-e)$ as a function of $a$, where $a$ and $e$ are the stars' semi-major axis and eccentricity, respectively. A star lying in this region of parameter space will inspiral due to GW radiation along curves like the solid red lines, and would take approximately the period of time labeled along the blue dashed lines to complete the inspiral, assuming no other interactions. Both are calculated using the approximate expressions in \citet{Peters:64}. A star lying below the solid grey line would likely turn into an EMRI, while one above would likely undergo some stellar scattering event which would increase its angular momentum (moving it upwards in the figure) and ultimately put it on a new trajectory with a much longer inspiral time. For reference and illustrative purposes we also plot the initial conditions of the $10^6$ stars we simulate. These stars are also appropriate for the standard EMRI formation, since they are drawn from a stellar distribution (\eqn{eqn:5:etarho}) in the absence of the secondary. For these highly eccentric orbits, in the standard EMRI picture, weak stellar scatterings extract or add angular momentum from a star while keeping its energy approximately constant. This moves the star up or down in the figure. When the star scatters below the gray line, gravitational radiation can extract energy from the orbit faster than it is likely to be perturbed by star-star scatterings, thus leaving the star to inspiral along the red lines. 

In the case presented in this paper, the picture is similar, but more intricate and rich. Kozai oscillations, along with significant apsidal precession due to the stellar potential and general relativity (GR), drive the stars on an orbital evolution which also conserves the orbital energy while causing the angular momentum to rise and fall before a close passage with the primary SMBH occurs. If the star passes close enough to the primary SMBH that a significant fraction of the star's orbital energy can be radiated in GWs, it can quickly circularise and form an EMRI. To elucidate these points the path of an EMRI formed in one of our simulations is shown in thin solid brown. The same star is also marked by a red-outlined gold star in \figlistthree{fig:5:kozaijc}{fig:5:kozai}{fig:5:oom} and used to demonstrate the relative importance of the stellar potential and GR in \fig{fig:5:revkozaiprecession}. Like many of the EMRIs formed in our simulations this star begins with a low eccentricity orbit and with a semi-major axis about a factor of $10$ smaller than the stalling radius of the secondary SMBH. It is only through an intricate interplay between the Kozai effect, precession due to the stellar potential (SP precession) and GR effects that the star is driven to high eccentricity, has strong interactions with the primary SMBH and ultimately forms an EMRI. 

%----------------------------------------------------------------------------
\subsection{Nomenclature and Notation}
\label{sec:5:notation}
%----------------------------------------------------------------------------
In general, we write quantities relevant to the primary SMBH without subscripts, those relevant to the secondary SMBH with a subscripted large black ``dot'', and those relevant to the stars with a subscripted star symbol. For example, the mass of the primary SMBH hole is written $M$, while those of the stars and secondary are written $\mstar$ and $\msec$, respectively. Similarly the semi-major axes of any given star and the secondary are written $\astar$ and $\asec$. Note, however, that the subscript is foregone in figure labels. The mass ratio of the two SMBHs is denoted $q\equiv \msec/\mprim\le1$. We summarise the notation used in this work in table \ref{tab:1:parameters}.

Because of the non-Keplerian potential (due to the stellar potential, the secondary SMBH and GR) there can be ambiguity when referring to the Keplerian orbital elements. We use unambiguous analogous quantities to the Keplerian orbital elements to describe the orbits of our stars. A star's semi-major axis, \astar, is defined to be a function of the star's energy, $E$,  
\BE
\astar \equiv {G\mprim\over 2E} \,, \label{eqn:5:aofE}
\EE
while the eccentricity is defined using the reduced angular momentum, \Ltot, and \astar:
\BE
e\equiv\sqrt{1-{\Ltot^2\over G\mprim a}} \,. \label{eqn:5:eofLa}
\EE
In these expressions $E$ and \Ltot are calculated in the primary SMBH's frame rather than the center of mass frame. These are the quantities shown in the figures unless stated otherwise. 

Since in static non-Keplerian potentials \Ltot and $E$ are conserved along orbits, then, provided the secondary's orbital period is long compared to a given star's, on the star's orbital timescale both $a$ and $e$ defined as above are well defined quantities and have minimal variations. Moreover, they are consistent with the standard definitions for the Keplerian orbital elements in the limit of Keplerian orbits.

The orbits which are interesting to us are those which approach the inner several Schwarzschild radii but have semi-major axes relatively close to the secondary SMBH; i.e., these orbits are highly eccentric. In this case, 
\BE
\Ltot^2\approx 2G\mprim a(1-e) \propto a(1-e) \,. \label{eqn:5:Lsquared}
\EE
The right hand side of this equation is  just the equation for the Keplerian periapsis distance in the limit of a Keplerian orbit. Because of this correspondence to both the angular momentum in these high eccentricity orbits and its correspondence to the periapsis distance in the Keplerian limit, we frequently plot $a(1-e)$ and refer to it as the periapsis distance.

Stars whose angular momentum is less than $\Lplunge\equiv 4GM/c$ on a close approach to a SMBH find themselves on a one-way trip to the black hole, `plunging' across the event horizon. This corresponds to a periapsis distance in Schwarzschild radial coordinate of $4GM/c^2$ (see \eqn{eqn:5:trp}), however in  $a(1-e)$, \Lplunge corresponds to $8GM/c^2$.

\begin{table*}
	\caption{\label{tab:1:parameters} Notation: Equations and Descriptions of Parameters and Variables Listed in Alphabetical Order}

	\begin{center}
	\begin{tabular}{|ccp{8.5cm}cl|}
	\hline\hline 
	Parameter && Description && Eqn.~\# \\
	\hline\hline
		$\astar$	&& 	semi-major axis of a star			&& \eqn{eqn:5:aofE} \\

	$\asec$ && 	semi-major axis of secondary SMBH && n/a \\
	$c$ 		&& 	speed of light 					&& n/a \\
	$e$ 		&& 	eccentricity 					&& \eqn{eqn:5:eofLa} \\
	$E$ 		&& 	specific orbital energy 					&& \sect{sec:5:notation} \\
	$G$ 		&& 	gravitational constant 			&& n/a \\
	$i$ 		&& 	inclination 			&& n/a \\
	\Ltot 		&& 	total specific angular momentum	&& \eqn{eqn:5:Lsquared} \\
	\Lz && $z$ component of specific angular momentum, SMBH binary lies in $x$-$y$ plane
	&& n/a \\
	\Lplunge 	&&	Angular momentum below which stars plunge into SMBH, equal to $4GM/c$ 
	&& n/a \\
	$\mprim$ 		&& 	mass of primary SMBH			&& n/a \\
	$\Mstar(<r)$&& stellar mass interior to $r$&& \eqn{eqn:5:massinterior} \\
	$\msec$ && 	mass of secondary SMBH	&& n/a \\
	$\mstar$ && 	mass of secondary SMBH		&& n/a \\
	$\pstar$ && 	period of star's orbit				&& n/a \\
	$\psec$ && 	period of secondary SMBH's orbit	&& n/a \\
	$q$		&& 	mass ratio of secondary to primary SMBH	&& n/a \\
	$r$ 		&& 	radial position from primary SMBH 	&& \sect{sec:5:stellardistribution} \\
	$r_c$ 	&& 	characteristic size of cusp 		&& \sect{sec:5:stellardistribution} \\
	$\rstall$ 	&& 	stalling radius of secondary 		&& \sect{sec:5:secondaryinspiral} \\
	$\tKoz$ 	&& 	Kozai timescale			 		&& \eqn{eqn:5:Tkozai} \\
	$\tkoz$ 	&& 	instantaneous Kozai period			 		&& \eqn{eqn:5:instantaneoustkozai} \\
	$\tphigr$ && 	time for orbit to precess by $\pi$ radians due to GR	&& \sect{sec:5:apsidalprecession} \\
	$\tphisp$ && 	time for orbit to precess by $\pi$ radians due to stellar potential	&& \sect{sec:5:apsidalprecession} \\
	$U(r)$ 	&& 	total gravitational potential		&& \eqn{eqn:5:pseudoNewtpotential} \\
	$v$ 	&& 	velocity		&& n/a \\
	
	\hline
	$\Delta \Lbin$ && oscillations in \Lz on the timescale of the secondary SMBH's orbit && \sect{sec:5:fluctuationsinLzbin}\\
	$\Delta \Lstar$ && oscillations in \Lz on the timescale of the stellar orbit && \sect{sec:5:fluctuationsinLzstar}\\
	$\eta$ 	&& 	parameter for cusp steepness 		&& \sect{sec:5:stellardistribution} \\
	$\rhostar$	&& 	density of stars			 		&& \eqn{eqn:5:etarho}\\
	$\rhostar(<v)$	&& 	density of stars with velocity less than $v$			 		&& n/a\\

	\hline
	\end{tabular}
	\end{center}
\end{table*}

%=============================================
\section{Physical Setup}
\label{sec:5:physicalsetup}
%=============================================

%----------------------------------------------------------------------------
\subsection{The Outline}
\label{sec:5:outline}
%----------------------------------------------------------------------------
Throughout this paper, the system we are considering is made up of three objects: 1) a primary SMBH surrounded by 2) a stellar cusp of mass equal to twice the primary's mass and 3) orbited by a secondary SMBH. We simulate this system by assuming that the stars in the cusp are non-interacting, allowing us to reduce the problem to a series of three-body problems which are made up of the primary SMBH, the secondary SMBH, and a star selected randomly from a stellar distribution.

In particular, we use a modified version of the simulation code used to study tidal disruptions in \cite[][hereafter \tdpaper, see \sect{sec:5:simulation} for differences]{Wegg:Bode:10}. There we were interested in the possibility of observing multiple tidal disruptions from the same galaxy due to the presence of a secondary SMBH. The similarities to the problem considered here make this code particularly appropriate.

We initially distribute stars isotropically according to an $\eta$-model \citep{Tremaine+94} for a single-mass stellar distribution around the primary SMBH (\sect{sec:5:stellardistribution}). However, we truncate the stellar potential just inside the stalling radius of the secondary (\sect{sec:5:fixedstellardistribution}). The secondary is then spiraled inwards on a slightly eccentric orbit approximating the orbit of a SMBH evolving by dynamical friction and stellar ejection (\sect{sec:5:secondaryinspiral}), but smoothly stopping the inward motion at the stalling radius \citep{Sesana+08b}. The primary difference between \tdpaper and here is that now we must take into account relativistic effects (\sect{sec:5:greffects}) to properly model the stellar dynamics.

%----------------------------------------------------------------------------
\subsection{The Initial Conditions}
\label{sec:5:initialconditions}
%----------------------------------------------------------------------------
We run four simulations, where we have varied two parameters: the SMBH mass ratio which is chosen to be either $q=0.1$ or $0.3$ and the stellar mass which is chosen to be either $1\msun$ or $10\msun$. The simulations are not independent, since we use the same initial velocities and positions for the $10^6$ stars simulated in all of the simulations. This is significantly quicker computationally because, for simulations with the same $q$, we only reintegrate orbits that pass within $100\,G\mprim/c^2$, the only region the stars' mass impacts its trajectory (which is only important when considering GR effects: see \sect{sec:5:gravitationalwavelosses}). Additionally, it also provides a direct comparison between the stars from each simulation which form EMRIs. 

We choose a primary SMBH mass of $M=10^6\,\msun$ for all of our simulations since this will result in EMRIs with frequencies best suited for detection by low-frequency space-based gravitational wave detectors such as LISA \citep{AmaroSeoane+07} or eLISA \citep{AmaroSeoane+12}. The $10^6$ stars are given initial positions and velocities appropriate for a relaxed isotropic cusp centred on the primary SMBH (\sect{sec:5:stellardistribution}).

%----------------------------------------------------------------------------
\subsection{The Stellar Distribution}
\label{sec:5:stellardistribution}
%----------------------------------------------------------------------------
We integrate stars drawn from a cusp centred on the primary SMBH. 
The initial orbits of the stars are drawn from the self-consistent isotropic potential-density pair known as the $\eta$-model, meaning that the initial stellar distribution is drawn from \citep[][with $\mu=0.5$]{Tremaine+94}
\BE
\rhostar(r)=\frac{\eta}{2\pi r_c^3}\frac{M}{\left(\frac{r}{r_c}\right)^{3-\eta} \left(1+\frac{r}{r_c}\right)^{1+\eta}} \, ,
\label{eqn:5:etarho}
\EE
where $r_c$ is the characteristic size of the cusp and $\eta$ is a dimensionless parameter controlling the central steepness of the cusp. The stellar mass interior to radius $r$ is therefore given by:
\BE
\Mstar(<r)\equiv M_{\star,\eta}(<r)={2 \mprim r^\eta\over(r_c+r)^\eta} \,. \label{eqn:5:massinterior}
\EE

Throughout this work we use $\eta=1.25$ since this is the relaxed form of the distribution close to the SMBH \citep{Bahcall:Wolf:76}. Our choice is motivated by \citet{Freitag+06} who simulate multi-mass models of stellar cusps finding mass segregation close to the black hole. In that work the most massive species are steeper than $\eta=1.25$ and less massive species less steep. However the overall density is close to $\eta=1.25$.

We discuss the consequences of our assumption of a universal $\eta=1.25$ in \sect{sec:5:discussion} and of mass segregation when when calculating the rates in \sect{sec:5:mergerprobability}.

%----------------------------------------------------------------------------
\subsection{Stellar Distribution: Static, Fixed and Truncated}
\label{sec:5:fixedstellardistribution}
%----------------------------------------------------------------------------
To reduce computational complexity we fix the stellar potential to the primary SMBH and do not evolve it in time. Though this is clearly inconsistent, since the secondary SMBH scatters stars onto new orbits and therefore modifies the stellar potential, most of our EMRIs are sourced from  a tenth of the stalling radius where these modifications are not significant. We discuss this in more detail in \sect{sec:5:assumptions} and demonstrate the magnitude of the error in \fig{fig:5:mintr}.

However, without modification there would be a different inconsistency of much greater import. It is important that the pre-calculated SMBH path is consistent with the integrated test particle equations of motion. If an inconsistency is present then a particle orbiting close to the primary has an incorrect acceleration towards the secondary. This would represent an unphysical dipole-like perturbation, which would typically dwarf the smaller tidal quadrupole perturbation due to the secondary. An inconsistency of this type would arise if \eqn{eqn:5:massinterior} was used as the stellar potential fixed to the primary, while the secondary orbit was calculated using \eqn{eqn:5:secorbit}, where only the stellar mass interior to the secondary's orbit appears. 

Instead, the stellar potential is truncated marginally inside the stalling radius. This has the result that the potential and stellar density are no longer self-consistent outside the stalling radius. However, stars outside the stalling radius that closely approach the SMBHs have undergone strong chaotic interactions with the binary. Therefore having a fully consistent stellar potential is less important in this region than for stars close to the primary which undergo secular interactions. For these inner stars we have the correct potential, and therefore the correct secular evolution.

%----------------------------------------------------------------------------
\subsection{Parameters of the Stellar Cusp}
\label{sec:5:stellarcusp}
%----------------------------------------------------------------------------

Throughout we use a fiducial cusp radius $r_c=1.7\pc$. This is motivated by the fits from \cite{Merritt+09} to the inner regions of ACS Virgo Cluster galaxies \citep{Cote+04}. For power-law galaxies these give\footnote{D. Merritt, personal communication. From fitting to \fignum{2} of \citet{Merritt+09}} 
\BE
r_{\rm inf} = 22\,(\mprim / 10^8 \msun)^{0.55}\pc \, , 
\label{eqn:5:rbullet} 
\EE
where $r_{\rm inf}$ is defined such that the stellar mass interior to $r_{\rm inf}$ is $2\, M$, and $M$ is the mass of the SMBH. Matching this to the $\eta$-model such that the central densities are equal gives $r_c=r_{\rm inf}$. Extrapolating to Sgr A* which has a mass of $\approx 4\times 10^6\,\msun$ \citep{Ghez+08} gives $r_{\rm inf}=3.8\pc$ which agrees well with the observations of $r_{\rm inf} \approx 4\, \pc$ \citep{Alexander:05}. Using $M=10^6\,\msun$ gives our fiducial $r_{\rm inf}=r_c=1.7\,\pc$. 

We note that using a total stellar mass which is twice that of the primary SMBH has the convenient property that when matching the central density to a power law, $r_c$ happens to be the radius at which the mass enclosed by the power law is $2\,M$ (the total stellar mass). This allows easy comparison with measurements.

%----------------------------------------------------------------------------
\subsection{The Inspiral of the Secondary}
\label{sec:5:secondaryinspiral}
%----------------------------------------------------------------------------
Initially the stars are on orbits consistent with the primary SMBH and the stellar potential. Subsequently their orbits are perturbed by interaction with the secondary SMBH as it inspirals to its stalling radius. In a fully self-consistent simulation the orbit of the secondary would evolve due to this exchange of energy with the stars. Instead, for efficiency and simplicity we calculate the orbit of the secondary SMBH assuming an inspiral dominated by dynamical friction with an appropriate Coulomb logarithm such that it stalls at the stalling radius. 

Specifically, the secondary SMBH is, at time $t=0$, given an eccentricity of approximately $0.1$ and an initial separation equal to the cusp radius, $r_c$. It is then migrated inwards on a path governed by
\BE
\frac{d \v}{dt} = -  \frac{G \left[\mprim (1+q)+\Mstar(<r) \right]}{r^3}\r - \frac{\v}{t_{\rm df}} \,
\label{eqn:5:secorbit} 
\EE
where $\Mstar(<r)$ is the stellar mass interior to $r$ and
\BE 
t_{\rm df}= \frac{v^3}{2\pi G^2\log\!\Lambda\, \mprim q \: \rhostar(< v)}
\EE
characterises the dynamical friction \citep{Binney:Tremaine:08}. Here $\rhostar(<\!v)$ is the density of stars at $r$ with velocity less than $v$. We have used a Coulomb logarithm that begins at $\log \Lambda\approx4$, but which smoothly decreases to zero at the stalling radius calculated by \cite{Sesana+08}. The functional form of the decrease was chosen to approximate the rate of shrinkage caused by the energy exchange with the stars during our simulations. This approximation was checked in \tdpaper.

All the simulations in this work are all of length $45 \sqrt{r_c^3/GM} \approx 1.5\Myr$. We assume that the secondary SMBH remains near it's stalling radius for this duration. Our choice of simulation termination is arbitrary and was chosen to limit computation time: the rate of EMRIs and direct plunges has significantly dropped but not yet fallen to zero at the end of the simulations. We repeated one of the simulations for four times the duration to assuage fears that only a small fraction of the total number of EMRIs were captured. The number of events in this longer simulation suggested we have captured approximately $\sfrac{2}{3}$ of the total \citep[for details see the more verbose][]{Wegg:Thesis}.

%----------------------------------------------------------------------------
\subsection{General Relativistic Effects}
\label{sec:5:greffects}
%----------------------------------------------------------------------------
There are two important GR effects which must be accounted for: gravitational wave energy/angular-momentum losses and periapsis precession. Ideally one would either integrate the exact equations of motion in full GR or use the simpler and well established post-Newtonian approximations. However, it is numerically prohibitive to attempt to use full GR, and the post-Newtonian expansions can be complex, are not separable, and do not have the appropriate divergences at low angular momenta. For example in the 3PN test particle limit the angular momentum at which parabolic orbits `whirl' (i.e. precession diverges) around a Schwarzschild black hole is $L \approx 4.69\,GM/c$ (compared to the correct value of $L = 4\,GM/c$) and unphysical bound orbits can occur below this \citep{Grossman:09}.

We instead choose the more straight forward approach of dealing with gravitational wave energy/angular-momentum losses and periapsis precession separately in our symplectic integrator.

Recently, \cite{wegg:12} proposed several pseudo-Newtonian potentials appropriate for the periapsis precession of test particles whose apoapses lies well beyond the Schwarzschild radius. We use  potential B of     that work, which balances both accuracy and computational cost.

As mentioned above, we deal with orbital energy and angular momentum losses due to gravitational wave emission separately. We deal with both by assuming the stars are on parabolic orbits. These losses are documented and fitting functions for the energy and angular momentum provided by \cite{Gair+05}. We use these fitting functions to determine the change in angular momentum and orbital energy during a complete periapsis passage, and apply these losses discretely at periapsis. The exact implementation is described in \sect{sec:5:gravitationalwavelosses}.

We note that for stars around Schwarzschild holes our methods remain accurate to arbitrary eccentricity (outside the exponentially narrow `whirling` region just above $L=4 \,G\mprim /c$ where the precession will be underestimated). The simulations are terminated only if the angular momentum drops below $\Ltot\le\Lplunge= 4 \,G\mprim /c$ as described in \sect{sec:5:selectionandrejection}.

%=============================================
\section{The Simulation}
\label{sec:5:simulation}
%=============================================
 
To integrate the orbits of the test particles we utilise the symplectic integrator described in \cite{Preto:Tremaine:99} and used in \cite{Peter:08,Peter:09I}. We use this integrator because
\BEn
\item Its symplectic nature causes energy to be conserved up to round-off error. This is desirable since spurious energy drifts would, over the many orbits simulated here, directly change the semi-major axis.
\item With an appropriate choice of step size (see \sect{sec:5:stepsize}) orbits in a Keplerian potential are reproduced exactly with only a phase error which is $\OO{N^{-2}}$.
\EEn

In particular, we use the version of the integrator used by \tdpaper, extended to take into account both relativistic precession (\sect{sec:5:pseudonewtonianpotential}) and the angular momentum and energy losses due to gravitational wave radiation (\sect{sec:5:gravitationalwavelosses}). 

For a detailed discussion of the integrator used in \tdpaper see that work, and the works on which it was based \citep{Preto:Tremaine:99,Peter:08,Peter:09I}. Here we discuss only how the integrator we use differs from that of \tdpaper, along with checks (\sect{sec:5:stepsize}) and selection criteria (\sect{sec:5:selectionandrejection}) relevant to this new context.

%-----------------------------------------------------------------------------
\subsection{Pseudo-Newtonian potential}
\label{sec:5:pseudonewtonianpotential}
%-----------------------------------------------------------------------------

The prime reason we must use a pseudo-Newtonian potential to model relativistic precession is that the symplectic integrator is constructed by operator splitting and hence requires the Hamiltonian be separable. We therefore cannot evolve particles using post-Newtonian approximations in the equations of motion \citep[as in e.g.][]{Merritt+11b}. 

Instead, relativistic precession is included using the Pseudo-Newtonian potential labeled as `potential B' in \citet{wegg:12}. We use this potential since it accurately reproduces the precession of orbits with apoapsis in the far field (i.e. apoapsis $\gg G\mprim/c^2$). Note that the potential of \citet{Paczynski+80} does not have this property. In addition, the precession correctly diverges as the orbit approaches \Lplunge which separates bound orbits from plunges (this occurs at $\Ltot=4G\mprim/c^2$ for parabolic orbits). For reference the complete potential used is
\BEA
U(r) &=& -\frac{G\mprim}{r_1} \left({1\over 1-{5\over6}{r_{\rm s,1}\over r_1}} +{2\over3}{r_{\rm s,1}\over{r_1}} \right)  \nn\\
&&\hspace{5mm}-\frac{G\msec}{r_2} \left({1\over 1-{5\over6}{r_{\rm s,2}\over r_2}} +{2\over3}{r_{\rm s,2}\over{r_2}} \right) - V(r_{1})\, ,\label{eqn:5:pseudoNewtpotential}
\EEA
where the numerical subscripts $1$ and $2$ are used to distinguish quantities measured with respect to the primary and secondary, respectively, $r_i$ is the distance to the $i$th SMBH, $r_{\rm s,i}$ is the Schwarzschild radius of the $i$th SMBH, and $V(r_1)$ is the stellar potential produced by \eqn{eqn:5:etarho}.

%-----------------------------------------------------------------------------
\subsection{Gravitational Wave Losses}
\label{sec:5:gravitationalwavelosses}
%-----------------------------------------------------------------------------

When an object passes close to either SMBH, relativistic effects such as energy and angular momentum losses due to gravitational radiation become important. We incorporate these changes into the orbit by stepping out of the symplectic integrator at periapsis and calculating a new velocity vector when the star has passed within $100\, GM/c^2$ of either SMBH. We note the calculations here are for non-spinning Schwarzschild black holes, this assumption is briefly discussed in \sect{sec:5:assumptions}.

The choice of $100\, GM/c^2$ is motivated by the low losses for stars that remain outside this radius: For our highest mass ratio ($10\msun/10^6\msun=10^{-5}$) the loss of angular momentum per orbit for a parabolic orbit with periapsis $100\, GM/c$ is $1.9 \times 10^{-8} GM/c$ and of energy is $1.6 \times 10^{-11} c^2$. For computational reasons (described below) the highest number of orbits integrated is $5\times10^4$, and therefore losses for orbits which remain outside of $100\, GM/c^2$ are negligible.

To compute the energy and angular momentum lost during a periapsis passage we assume the orbits are parabolic. We then relate the angular momentum in the orbit to the energy and angular momentum lost during each periapsis passage using the fitting functions for parabolic orbits from \cite{Gair+06}.  In that paper they compute fitting functions to the energy and angular momentum loss using the Teukolsky equation. 
For convenience we provide these fitting functions here:
\BEA
{M\over m}\Delta X&=&\cosh^{-1}\left[1+B_0^X\left({4\over \trp}\right)^{N_X-1}{1\over \trp-4}\right]\nn\\
&&\times\sum\limits_{n=0}^NA_n^X\left({1\over \trp}-{4\over \trp^2}\right)^n\nn\\
&&+{\trp-4\over \trp^{1+N_X/2}}\sum\limits_{n=0}^NC_n^X\left({\trp-4\over \trp^2}\right)^n\nn\\
&&+{\trp-4\over \trp^{2+N_X/2}}\sum\limits_{n=0}^{N-1}B_{n+1}^X\left({\trp-4\over \trp^2}\right)^n \,,\label{eqn:5:fittingfunction}
\EEA
where $X$ is either the specific energy $E/c^2$ or the (scaled) specific angular momentum $\tL=\Ltot/(G\mprim/c)$, $\trp$ is the periapsis distance in geometrized units, $N_E=7$, $N_L=4$, and the $A_n^X$, $B_n^X$, and $C_n^X$ are coefficients given in \tab{tab:5:fittingfunction}. In \cite{Gair+06} they note that $N=2$ is sufficient for better than $0.2\%$ accuracy everywhere. This is the order used in our code. Here, $\trp$ is calculated based on the periapsis an orbit would have if it were parabolic and had the measured angular momentum:
\BE
\trp = \frac{\tL^2}{4} \left( 1 + \sqrt{1-\frac{16}{\tL^2}} \right) \, . \label{eqn:5:trp}
\EE

Note that $\trp$ is {\it not} calculated from the position of the star output by our simulation at its ostensible periapsis, since in the pseudo-Newtonian potential this does not match its relativistic value. 

\begin{table}
\begin{center}
\caption{Coefficients for \eqn{eqn:5:fittingfunction} \label{tab:5:fittingfunction}}
\begin{tabular}{|c|ccc|}
\hline
\hline
&$n=0$&$n=1$&$n=2$\\
\hline
\hline
$A_n^E$	&$-0.318434$	&$-5.08198$	&$-185.48$\\
$B_n^E$	&$0.458227$	&$1645.79$	&$8755.59$\\
$C_n^E$	&$3.77465$	&$-1293.27$	&$-2453.55$\\
\hline
$A_n^L$	&$-2.53212$	&$-37.6027$	&$-1268.49$\\
$B_n^L$	&$0.671436$	&$1755.51$	&$9349.29$\\
$C_n^L$	&$4.62465$	&$-1351.44$	&$-2899.02$\\
\hline
\end{tabular}
\end{center}
\end{table} 

We subtract the energy and angular momentum loss given by \eqn{eqn:5:fittingfunction} at the step closest to periapsis. 
At this step we calculate a new velocity, $\vec{v'}$, using the new specific energy, $E'=E+\Delta E$, and specific angular momentum, $\Ltot'=\Ltot+\Delta \Ltot$. Since the position is unchanged, the potential energy is unchanged and 
\BE 
v'^2 = v^2 + 2 \Delta E \label{eq:deltav2}\,.
\EE 
The orbital plane remains unchanged for a Schwarzschild black hole and therefore
\BE 
\Lvec'=\frac{\Ltot+\Delta \Ltot}{\Ltot}\vec{\Ltot}=\vec{r}\times\vec{v'} \, . \label{eq:hprime}
\EE
Taking the dot product of this yields
\BE 
\vec{r}.\vec{v'}=\sqrt{\Ltot'^2 - r^2 v'^2} \, \label{eq:rdotv}
\EE
where we take the positive branch of $\vec{r}.\vec{v'}$ since this corresponds to the outgoing, post-periapsis solution. The cross-product $\vec{r}\times\Lvec'$ yields
\BE 
\vec{v'} = \frac{1}{r^2} \left[ (\vec{r}.\vec{v'}) \vec{r} - \vec{r}\times\Lvec' \right] \label{eq:vprime} \, .
\EE
\Eqn{eq:vprime} together with \ref{eq:deltav2}, \ref{eq:hprime}, and \ref{eq:rdotv} are then used to calculate the new velocity $\vec{v'}$ following the periapsis passage. 

\begin{figure}
\centering
\includegraphics{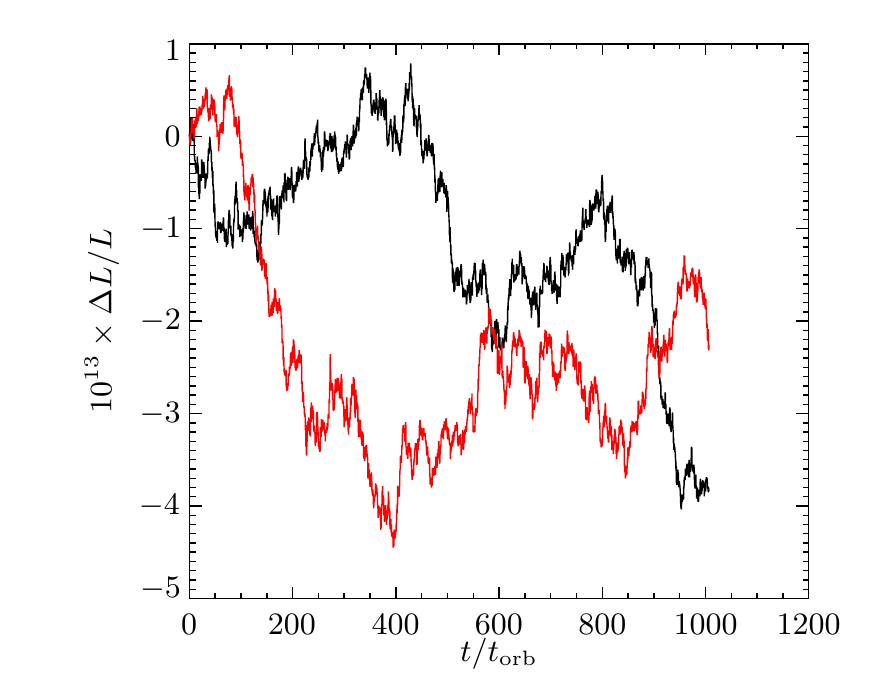}
\caption[Plot showing the errors in conservation of reduced angular momentum]{Plot showing the errors in conservation of reduced angular momentum, \Ltot, over many orbital periods ($t_{\rm orb}$) of a high eccentricity ($e=1-10^{-5}$) test particle.  The red is without the procedure for calculating the change in $\vec{v}$ at periapsis, while the black uses \eqn{eq:vprime} but with $\Delta \Ltot=0$ . The secondary has zero mass for both curves. The errors are  still at the level $\Delta \Ltot/\Ltot\sim10^{-12}$ indicating that the process of stepping in and out of the symplectic integrator does not inherently introduce significant errors.}
\label{fig:5:hcons}  
\end{figure}

In \fig{fig:5:hcons} we show the numerical accuracy of this procedure by considering whether \Ltot remains constant over many orbits. 

%-----------------------------------------------------------------------------
\subsection{Step Size}
\label{sec:5:stepsize}
%-----------------------------------------------------------------------------

\citet{Preto:Tremaine:99} show that, for their adaptive symplectic integrator, in a Keplerian potential $U\propto 1/r$ then using a step size $\propto 1/r$  reproduces Keplerian orbits exactly with a phase error whose size is $\OO{N^{-2}}$, where N is the number of steps per orbit. Since the stars considered here usually evolve in a nearly Keplerian potential we therefore use a step size $\propto U$. The method of choosing the step size has not changed from \tdpaper, except that here we have chosen to have $20,000$ steps per orbit.

The phase error introduced by the finite step size in the integrator is along the orbit in a Keplerian potential i.e. a time error. The dominant error is also along the orbit for the pseudo-Newtonian potential. We expect both errors to be unimportant to both the dynamics and the rates, provided they are small per orbit.

Estimating both phase errors per orbit:
\begin{inparaenum}
\item The phase error from using a finite step size in the integrator is of order $1/N^2 \sim 10^{-8}$.
\item The phase error in using our pseudo-Newtonian potential compared to the geodesic equation is of order $E/c^2 = GM/ac^2$. Our EMRIs are sourced from the dynamics which occur at $a \sim 0.007\pc$. The phase error for these stars is therefore of order $7\times10^{-6}$ per orbit.
\end{inparaenum}

These errors are also small per orbital period of the secondary SMBH. The pseudo-Newtonian potential accumulates a time error of order unity over the entire simulation at $a \sim 0.007 \pc$. However we do not expect this to affect the dynamics.

To check that we are not sensitive to step size we re-ran a simulation with $10,000$ steps per orbit. While individual stars evolved differently due to the chaotic nature of some orbits, the number of EMRIs and plunges were statistically unchanged.

%-----------------------------------------------------------------------------
\subsection{Selection and Rejection}
\label{sec:5:selectionandrejection}
%-----------------------------------------------------------------------------
Throughout the stars' orbits their periapsis distances and semi-major axes were monitored. The simulation of a given star was stopped if one of three criteria were met: if the star was deemed an EMRI, a plunge, or beyond our computational capacity.

A star was labeled as an EMRI and it's evolution terminated if it entered the (e)LISA band. We choose this to correspond to a semi-major axis where the test particle's orbital period is below $5000\s$, i.e. when the semi-major axis is less than 
\BE 
a^3 < GM \left( \frac{5000\s}{2\pi} \right)^2 \,.
\EE
The results are not sensitive to this choice of stopping criteria.

A star's simulation was also stopped if the star's angular momentum at periapsis was less than the plunge angular momentum ($\Ltot\le\Lplunge= 4 \,G\mprim /c$). In this case the star was labeled as a plunge.

In both cases the star was subsequently reintegrated without a secondary to check that it would not otherwise have become an EMRI or plunge. In particular stars that over our simulation with no secondary, would have lost more than 5\% of their energy to gravitational radiation, or have plunged into the primary hole are discarded and not included in the results.

Finally, due to computational limitations the stars' evolutions were limited to $10^{10}$ steps. This affects stars with semi-major axes less than $10^{-3}$ pc (see \figsand{fig:5:oom}{fig:5:mintr}). These stars are both theoretically (see \sect{sec:5:parameterspace}) and empirically (see \figsand{fig:5:kozaijc}{fig:5:kozai}) unlikely to form EMRIs.

%=============================================
\section{Results and EMRI Rates}
\label{sec:5:emrirates}
%=============================================

\begin{table*}
\begin{center}
\begin{minipage}[c]{\linewidth}
\centering
\caption{Number of EMRIs and Their Probabilities in the Simulations \label{tab:5:simulationresults}}
\begin{tabular}{|c|cccc|cc|}
\hline
\hline
&\multicolumn{4}{c|}{Parameters}&\multicolumn{2}{c|}{EMRI stars}\\
\hline
\multirow{2}{*}{Sim \#}	
&\multirow{2}{*}{$q$\footnote{$q=\msec/\mprim$ is the ratio of the mass of the secondary SMBH to the primary SMBH.}}
&\multirow{2}{*}{${\mstar\over\msun}$\footnote{The assumed mass of the stars in \msun. The mass of the stars is only relevant when the star passes within $100\,G\mprim /c^2$ of one of the SMBHs.}} 	
& \multirow{2}{*}{$N_\star$\footnote{The total number of stars simulated during the run.}}	
&\multirow{2}{*}{Duration (Myr)\footnote{The duration of the simulation in megayears.}} 
&\multirow{2}{*}{$N_{\rm EMRI}$\footnote{The total number of EMRIs formed during the simulation.}}
& \multirow{2}{*}{$P_{\rm EMRI}$\footnote{The probability of a CO of mass $\mstar$ becoming an EMRI.}} \\
&&&&&&\\
\hline
\hline
1	&$0.3$	&$10$	&$10^6$	&$1.5$	&$10$	&$1.0\times 10^{-5}$	\\
2	&$0.3$	&$1$	&$10^6$	&$1.5$	&$1$	&$0.1\times 10^{-5}$\\
3	&$0.1$	&$10$	&$10^6$	&$1.4$	&$13$	&$1.3\times 10^{-5}$	\\
4	&$0.1$	&$1$	&$10^6$	&$1.4$	&$3$	&$0.3\times 10^{-5}$	\\
\hline 
\end{tabular}
\end{minipage}
\end{center}
\end{table*}

%============================
\subsection{Introduction to Results and Rates}

We provide a summary of the results of our simulations in \tab{tab:5:simulationresults}, including both the total number of EMRIs produced and the implied probability of forming an EMRI given the simulation's parameters.

From \tab{tab:5:simulationresults} calculating the rates is straightforward: Given a species $X$ of CO (stellar mass black hole, neutron star, or white dwarf) the simulation simulates $N_\star$ test particles and outputs the number of EMRIs $N_{\rm EMRI}(X)$ assuming a mass $m_X$. Then 
\BE
P_{\rm EMRI}(X)\equiv N_{\rm EMRI}(X)/N_\star  \label{eqn:5:pemri}
\EE 
is the probability that a star of species $X$ eventually becomes an EMRI (\sect{sec:5:mergerprobability}). 

We then multiply by the expected number of  stars in our model cusp of species $X$ assuming some quantity of mass segregation (\sect{sec:5:speciesnumberdensity}). This gives the approximate number of events for a given galaxy during the period of time that a secondary SMBH is settling to its stalling radius. 

By determining the volumetric rate of galaxies undergoing a gasless merger, $\dot{n}_{\rm merger}(\mprim =10^6\,\msun)$ (\sect{sec:5:mergernumberdensity}), we may produce the predicted EMRI rate density (\sect{sec:5:finalrates}). That is,

\BE
\mathcal{R}_{\rm EMRI}(X)=P_{\rm EMRI}(X)N_X\dot{n}_{\rm merger}(\mprim =10^6\msun) \,. \label{eqn:5:rateequation}
\EE

%----------------------------------------------------------------------------
\subsection{EMRI Merger Probability}
\label{sec:5:mergerprobability}
%----------------------------------------------------------------------------

Each of our simulations has $N_\star=10^6$, with stars of mass $10\,\msun$ or $1\,\msun$ (but not both), where we use the former to predict the rates of stellar mass black holes and the latter to predict the rates for both neutron stars and white dwarfs. The probability that an object in a given simulation will turn into an EMRI is calculated using \eqn{eqn:5:pemri} and shown in the final column of \tab{tab:5:simulationresults}.

%----------------------------------------------------------------------------
\subsection{Species Number Density}
\label{sec:5:speciesnumberdensity}
%----------------------------------------------------------------------------

\begin{table*}
\begin{center}
\begin{minipage}[c]{\textwidth}
\centering
\caption{\label{tab:5:speciesdensities} Approximate Mass and Number Densities of Species Deep ($<0.05\pc$) in Stellar Cusp found by \citep{Freitag+06}.}
\begin{tabular}{|c|ccc|c|}
\hline
\hline
\multirow{2}{*}{Species $X$}	
&\multirow{2}{*}{$m_X\over\msun$\footnote{Mass of species $X$ in solar masses.}} 	
&\multirow{2}{*}{$\rho_X\over\rho_{\rm MSS}$\footnote{Ratio of density of species $X$ to that of main sequence stars in the region where EMRIs are formed in our simulations \citep{Freitag+06}.}}
& \multirow{2}{*}{$N_X\over N_{\rm MSS}$\footnote{Ratio of number of stars of species $X$ to the number of main sequence stars in the region where EMRIs are formed in our simulations.}}	
&\multirow{2}{*}{$N_X$\footnote{Total number of species $X$ in the entire stellar cusp if the cusp were to have the same ratio of species $X$ to main-sequences stars as in the region where the EMRIs are sourced in our simulation.}} \\
&&&&\\
\hline
\hline
Main Sequence Star	&$1$	&$1$		&$1$		&$2\times10^5$	\\
Stellar-Mass Black Hole	&$10$	&$\sim 10$	&$\sim1$		&$2\times10^5$  \\
Neutron Star			&$1$	&$\sim 0.1$	&$\sim0.1$	&$2\times10^4$	\\
White Dwarf			&$1$	&$\sim 0.3$	&$\sim0.3$	&$6\times10^{4}$	\\
\hline
\end{tabular}
\end{minipage}
\end{center}
\end{table*}

The number of compact objects of type $X$ expected as a function of position in the cusp is poorly understood, both observationally and theoretically
\citep{Hopman:06,OLeary:09,Alexander:09}. We note that mass segregation would not effect the migration of the binary we calculate in \sect{sec:5:secondaryinspiral} (although overall density profiles different from $\eta=1.25$ would). Since the details of mass segregation are uncertain, and most of the EMRIs initiate from a narrow range of semi-major axes, we choose a somewhat novel approach. When estimating the rates we assume the entire cusp at these radii is formed either of stellar mass BHs, or WDs and NSs. The reader can then scale the numbers to their preferred ratios of objects at $a \sim 0.01\pc$.  This is possible since in the case considered the rates scale linearly with number of COs. Our fiducial choice here is to scale to the degree of segregation found by \citet{Freitag+06}.
We tabulate these values in column $b$ of \tab{tab:5:speciesdensities} and use them as our fiducial values. 

To determine the number of each stellar remnant in the cusp we must first consider stellar-mass black holes. In \cite{Freitag+06} they find that there is roughly $10$ times as much mass in SBHs than in MSSs close to the SMBH. We assume here that SBH masses are $10\msun$, which then tells us that there are roughly the same number of SBHs as there are MSSs in the cusp. Because most of the mass in the cusp is from the SBHs we can approximate the total number of SBHs by dividing the total mass of the cusp, which we have assumed to be $2\times10^6\msun$, by the mass of the SBHs. This yields $2\times10^5$ stellar-massed black holes in the cusp. This then tells us there are roughly $2\times10^5$ MSSs. 

As with stellar-massed black holes \cite{Freitag+06} provides the mass density ratios of each stellar type to that of MSSs. Therefore, assuming that NSs ans WDs are roughly $1\msun$, it is trivial to determine the number of each species in the cusp relative to the number of MSSs. Multiplying by $N_{\rm MSS}$ then gives the total number of each species in the cusp: 
\BE
N=\begin{cases}
2\times10^5 & \text{for main sequence stars}\\
2\times10^5 & \text{for stellar-mass black holes}\\
2\times10^4 & \text{for neutron stars}\\
6\times10^4 & \text{for white dwarfs}
\end{cases} \ .
\EE
These calculations are summarized in column $d$ of \tab{tab:5:speciesdensities}.

The predicted number of stellar-massed BHs in \tab{tab:5:speciesdensities} implies an unrealistic total number of stellar-massed BHs in the cusp, but these numbers have been scaled to ensure the correct number of stellar-massed BHs in the inner cusp, from where the EMRIs are sourced. Thus, these numbers produce the correct number of stellar-massed BHs in the relevant region.

%----------------------------------------------------------------------------
\subsection{Number Density of Mergers}
\label{sec:5:mergernumberdensity}
%----------------------------------------------------------------------------

We approximate the number density of mergers of SMBHs of mass $10^6\,\msun$ and mass ratios between $0.1$ and $0.3$ by determining the number density of SMBHs of mass $(10^{5.5}-10^{6.5})\,\msun$ and assuming one such SMBH merger per galaxy lifetime.

\cite{Aller:Richstone:02} find a local number density of SMBHs with mass ($10^{5.5}$--$10^{6.5}$)$\,\msun$ of $4\times10^6 \text{ Gpc}^{-3}$ (assuming $H_0=70\,\km\,\s^{-1}{\rm Mpc}^{-1}$). Therefore we approximate,
\BE
\dot{n}_{\rm merger}(\mprim =10^6\msun)\sim3\times10^{-4} {\text{mergers}\over \text{ Gpc}^{3}\,\text{yr}} \,. \label{eqn:5:mergerrate}
\EE

We have tacitly assumed above that mergers happen uniformly in time.
The assumed constant merger rate (of one merger per galaxy) will probably be fairly accurate for a LISA or a LISA-like experiment able to detect EMRIs to redshift $z\sim 1$. However because of the locally falling merger rate they would be over estimates if, for example, EMRIs are detectable only to redshifts of $z\sim 0.1$. However there are large uncertainties both in the design of any LISA like mission, the detectability of EMRIs, and the cusps and the degree of their relaxation and segregation around $10^6 \msun$ SMBHs. The SMBH merger rates used here are chosen due to their simplicity given these uncertainties.

%----------------------------------------------------------------------------
\subsection{Final Rates}
\label{sec:5:finalrates} 
%----------------------------------------------------------------------------

The rate of EMRI production per unit volume, $\mathcal{R}_{\rm EMRI}$, is calculated using \eqn{eqn:5:rateequation}. We use the probability that each star in our simulations becomes an EMRI, $P_{\rm EMRI}$, from \tab{tab:5:simulationresults}, the numbers of each species, $N_X$, from \tab{tab:5:speciesdensities}, and the SMBH merger rate, $\dot{n}_{\rm merger}$, from \eqn{eqn:5:mergerrate}. 

This analysis ultimately yields the following rates
\BE
\mathcal{R}_{\rm EMRI}(q=0.1)=
\begin{cases}
    8\times10^{-4}\ \yr^{-1}\Gpc^{-3}& \text{for SBHs}\\
    2\times 10^{-5}\ \yr^{-1}\Gpc^{-3}& \text{for NSs}\\
    5\times10^{-5}\ \yr^{-1}\Gpc^{-3}& \text{for WDs}
\end{cases}
\EE
and
\BE
\mathcal{R}_{\rm EMRI}(q=0.3)=
\begin{cases}
    6\times10^{-4}\ \yr^{-1}\Gpc^{-3}& \text{for SBHs}\\
    6\times 10^{-6}\ \yr^{-1}\Gpc^{-3}& \text{for NSs}\\
    2\times10^{-5}\ \yr^{-1}\Gpc^{-3}& \text{for WDs}
\end{cases}\,.
\EE
For reference these rates are given in \tab{tab:5:finalrates}.

For context our overall detection rates are significantly lower than the $\mathcal{R}\sim 1 \yr^{-1}\Gpc^{-3}$ predicted from isolated SMBHs by \cite{Gair:04}. We note that there is presently considerable uncertainly in each rate estimate, and show in \sect{sec:5:emriratesdiscussion} that the rate given here has the prospect of being astrophysically interesting to LISA-like missions.

\begin{table*}
\begin{center}
\begin{minipage}[c]{\textwidth}
\centering
\caption{\label{tab:5:finalrates} Final Rates of EMRIs Due To SMBH Binaries  }
\begin{tabular}{|c|ccc|}
\hline
\hline
\multirow{3}{*}{$q$\footnote{The mass ratio of the secondary SMBH to the primary}}	
&\multicolumn{3}{c|}{\multirow{2}{*}{$\mathcal{R}_{\rm EMRI} \text{ (yr$^{-1}$Gpc$^{-3}$)}$\footnote{Rate of EMRIs due to SMBH binaries for the cases of stellar-mass black holes (SBHs), neutron stars (NSs), and white dwarfs (WDs) assuming that all mergers at rate $\dot{n}_{\rm merger}$ have mass ratio $q$}}}	\\
&&&\\\cline{2-4}
&SBH&NS&WD\\
\hline
\hline
$0.3$	&$6\times 10^{-4}$	&$6\times10^{-6}$	&$2\times10^{-5}$\\
$0.1$	&$8\times 10^{-4}$	&$2\times10^{-5}$	&$5\times10^{-5}$	\\
\hline
\end{tabular}
\end{minipage}
\end{center}
\end{table*}

%--------------------------------
\begin{figure*}
\begin{center}
\roughdraft{\includegraphics[width=0.9\textwidth]{./Figures/comp.png}}{\includegraphics[width=0.9\textwidth]{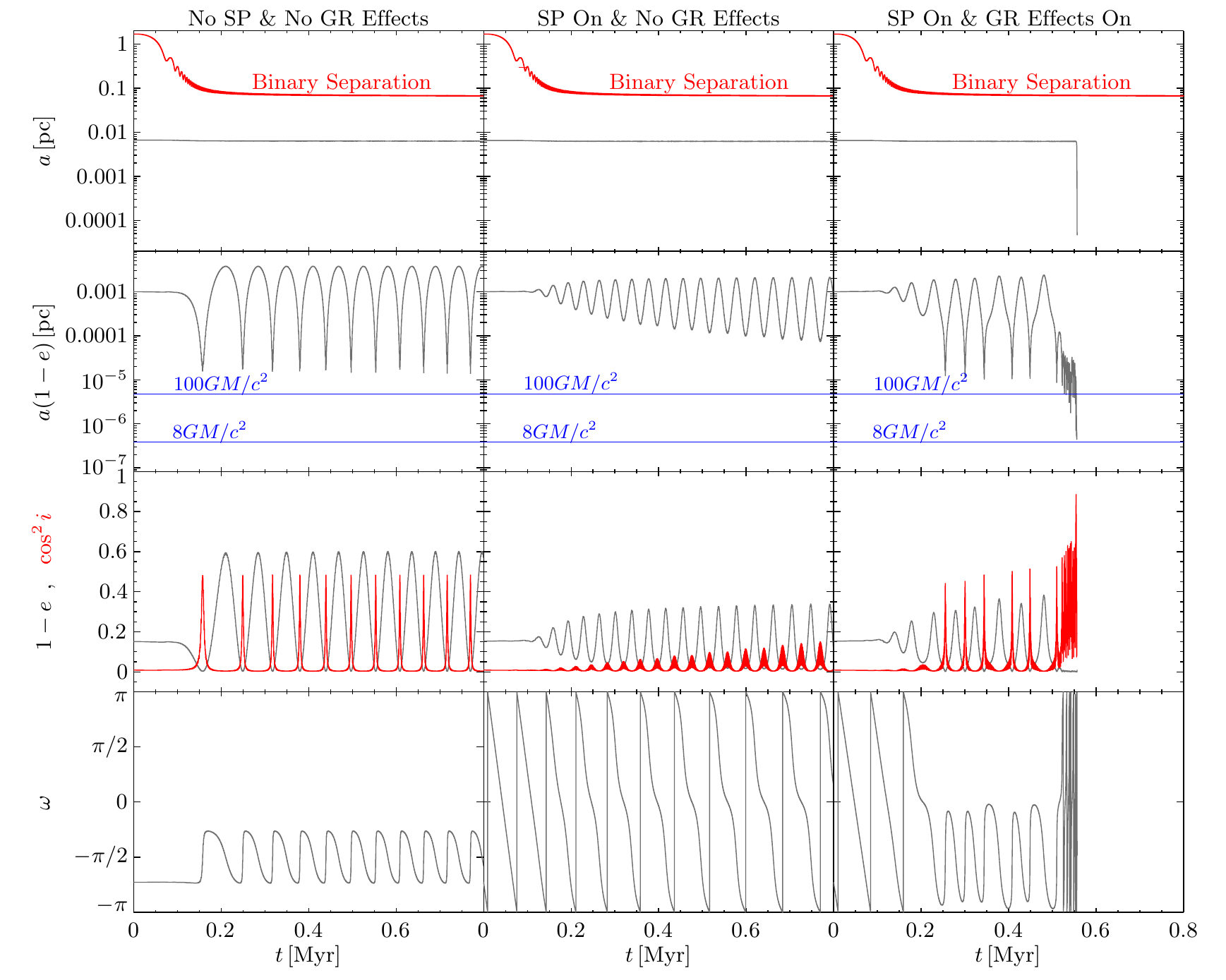}} 
\caption[Comparison of Kozai features for a specific star being integrated in our simulation with different forms of precession]{\label{fig:5:revkozaiprecession} Comparison of Kozai features for a star which turned into an EMRI in one of our simulations evolved with different forms of precession: in the first column we follow the star without any form of precession (no SP or GR precession), in the second column we ``turn on'' SP precession and in the third column we show the results of the full integration which also has GR precession and GW energy loss. Describing the rows of the plot: In row 1 we plot in charcoal the semi-major axis of a $10\,\msun$ star (the mass is only relevant when GR effects are considered; final column) as a function of time along with the \emph{position} of the secondary plotted in red. In row 2 we plot $a(1-e)$ in charcoal as a function of time. Also plotted are the lines of constant  $a(1-e)=8\,G\mprim /c^2$ and $100\,G\mprim /c^2$, the latter being the arbitrary radius where we start calculating the energy loss for the orbit (see \sect{sec:5:simulation}). In row 3 we plot $1-e$ in charcoal and $\cos^2i$ in red as a function of time. The product of the two is proportional to $\Lz^2$ which is a conserved quantity in the standard Kozai formalism. In the final row we plot the argument of periapsis as a function of time. From left to right there are clear changes in the star's orbit as new forms of precession are added: In the first column once the secondary has reached the stalling radius the star undergoes librating oscillations about $\omega=-\pi/2$ most akin to the traditional Kozai oscillations as described in \sect{sec:5:kozaiintro}. In the second column three clear changes occur: the star begins retrograde precession (row 4) as the SP precession dominates, the oscillation period decreases(rows 2, 3 and 4), and the magnitude of the oscillations is reduced (rows 2 and 3). In the final column (with SP and GR effects on) a very different phenomenon occurs which can be separated into three regions: strong retrograde precession due to the stellar potential, a librating mode in a pseudo-Kozai oscillation where SP and GR precession loosely cancel, and a final phase marked by rapid, and apparently chaotic, prograde precession driven by GR. Again, the period of oscillations in $e$ and $i$ are half the precession period. The increasing relevance of GR precession is due not only to having a lower periapsis distance, but also to the effects of the SP being weaker as the eccentricity grows. This star is also marked as a red-outlined gold star in \figlistthree{fig:5:kozaijc}{fig:5:kozai}{fig:5:oom}. }
\end{center}
\end{figure*}
%--------------------------------

%=============================================
\section{Stellar Dynamics: Understanding Our EMRIs and Plunges}
\label{sec:5:oom}
%=============================================

The processes which produce the majority of our EMRIs are, dynamically, quite rich. In particular, the interplay between the physical processes of the secular Kozai effect, the SP precession, the competing GR precession, and the oscillations of the orbital elements on the secondary SMBH's orbital timescale is physically intricate and interesting. We elucidate these effects in this section, where we first give a brief description of the Kozai effect (\sectsand{sec:5:kozaiintro}{sec:5:kozaiadditionalpoints}), then discuss various complicating effects individually (\sects{sec:5:apsidalprecession}{sec:5:fluctuationsinLzstar}), and finally consider the different effects together elucidating their relevance to the EMRI rates (\sect{sec:5:oom}). Of particular importance are the oscillations in the angular momentum on the timescale of the secondary (\sect{sec:5:fluctuationsinLzbin}).

%============================
\subsection{Kozai Effect --- Historical Formalism}
\label{sec:5:kozaiintro}
%============================

Instead of giving a detailed description of the Kozai mechanism, something already comprehensively elucidated by the original papers \citep{Lidov:62,Kozai:62} and much subsequent work \citep{Holman+97,Blaes+02,Ivanov+05,Thompson:10} in various different contexts, we aim to provide in this section the key equations and their consequences relevant to our problem.

In our circumstances the Kozai mechanism is a secular process whereby the weak quadrupolar tidal force from the secondary SMBH perturbs the orbits of stars around the primary SMBH.

The original theory \citep{Lidov:62,Kozai:62} of the Kozai mechanism assumed (in the context of our problem) not only that the semi-major axis of the star is significantly less than that of the secondary SMBH, but also that the star is on what would otherwise be a Keplerian orbit in the absence of the secondary (i.e., general relativistic effects along with effects due to the stellar potential are ignored). Moreover, the Kozai-Lidov theory assumes a purely quadrupolar force and averages over the orbits of both the star and the secondary SMBH.  Throughout this subsection we retain these assumptions.

Here, as with elsewhere in this paper, we will use the following conventions: $\omega$ is the argument of periapsis, $\chi$ is the longitude of the ascending node, $\psec$ is the period of the secondary, $\pstar$ is the radial period of the star, and 
\BE
\tKoz\equiv {2\over 3\pi q}{\psec\over \pstar}\psec ={4\over 3q}\left({\astar\over \asec}\right)^{-3/2}\sqrt{\asec^3\over G \mprim} \label{eqn:5:Tkozai}
\EE
is the widely discussed characteristic timescale on which the Kozai oscillations occur \citep[e.g.,][]{Ivanov+05}.

Starting from the exact equations of motion in the osculating elements, written with the true anomaly as the independent variable, and averaging over both the orbits of the star and the secondary SMBH, \cite{Lidov:62} obtained an insightful set of differential equations (written in the form of \citealt{Ivanov+05}):
\BEA
\tKoz {da\over dt} &=& 0 \label{eqn:5:dadt}\\
\tKoz {de\over dt} &=& -{5\over2}e\sqrt{1-e^2}\sin^2i\sin2\omega \label{eqn:5:dedt}\\
\tKoz{di\over dt} &=& -{5\over4}{e^2\sin(2i)\sin(2\omega)\over\sqrt{1-e^2}}\label{eqn:5:didt}\\
\tKoz{d\omega\over dt} &=&{2(1-e^2)+5\sin^2(\omega)(\cos^2i-(1-e^2))\over\sqrt{1-e^2}}\label{eqn:5:dwdt}\\
\tKoz{d\chi\over dt} &=& -{\cos i\over \sqrt{1-e^2}}\left\{1+e^2(5\sin^2\omega-1)\right\} \label{eqn:5:dchidt}\, .
\EEA

The primary characteristic of the Kozai mechanism is that the star's orbital elements undergo an oscillatory motion which has a period given approximately by $\tKoz$, and which can be of significant magnitude. There are several key outcomes from the above equations:
\BEn
\item $a$, and, therefore, the energy of the star's orbit remains constant (\eqn{eqn:5:dadt}).
\item \label{pt:5:eextrema} The eccentricity and inclination reach their extremal values only when $\omega=0$, $\pm\pi/2$, {\it or} $\pi$ (solving \eqn{eqn:5:dedt} equal to $0$).
\ifthenelse{\boolean{thesis}}{
\item Setting $\dot{e}=\dot{i}=\dot{\omega}=0$ we find a stationary solution when $\omega=\pm\pi/2$. In this case $e$ and $i$ are fixed and obey the relation
\BE
e^2={5\over3}\sin^2i-{2\over3} \label{eqn:5:libratingextrema}\,,
\EE
while $\omega=\pm{\pi/2}$.
}{}
\EEn

In addition to $a$, equations \ref{eqn:5:dadt}-\ref{eqn:5:dchidt} admit two further integrals of the motion \citep{Lidov:62,Kozai:62}:
\BEA
\Theta=(1-e^2)\cos^2i &\hspace{0.5cm}&\text{ is conserved and} \label{eqn:5:Theta}\\
Q=e^2\left[5\sin^2i\sin^2\omega-2\right]&\hspace{0.5cm}&\text{ is conserved} \label{eqn:5:Q} \,,
\EEA
which together tell us several things about a star's evolution:
\begin{enumerate}[resume,leftmargin=0.5cm]
\item \label{pt:5:Lzconserved}The $z$ component of the angular momentum is conserved since $\Lz = \sqrt{G \mprim a \Theta}$ (\eqn{eqn:5:Theta}).
\item All solutions have $1-e^2\ge\Theta$ or, equivalently, $e<\sqrt{1-\Theta}$ (\eqn{eqn:5:Theta}).
\item The eccentricity reaches its maximum (minimum) when the inclination reaches its minimum (maximum) (\eqn{eqn:5:Theta}).
\EEn

Many of these points can be seen in the first column of \fig{fig:5:revkozaiprecession}, where we have plotted the evolution of a star whose fate in our $q=0.3$ and $\mstar=10\,\msun$ simulation was to become an EMRI (also shown in  \figlistfour{fig:5:emriparameterspace}{fig:5:kozaijc}{fig:5:kozai}{fig:5:oom}), except with the stellar potential and GR effects turned off. While the semi-major axis of the star remains constant (first row), \Ltot goes through significant oscillations. To avoid confusion we reiterate that in our simulations we are \emph{not} solving \eqns{eqn:5:dadt}{eqn:5:dchidt}, but rather directly integrating the full three-body (non-orbit averaged) calculation (see \sect{sec:5:simulation} for detailed explanation).

%============================
\subsection{Conservation of \Lz in Standard Kozai-Lidov Formalism}
\label{sec:5:Lzconserved}
%============================

Point \ref{pt:5:Lzconserved} above has particular importance to our problem. Since \Lz is conserved, it must be true that $\Ltot\ge\Lz$ through out the evolution of a star. Because EMRIs, tidal disruptions, or plunges all require a low total angular momentum of order $\Lplunge$ or smaller, if a star is to be driven to be an EMRI, tidal disruption, or plunge by the Kozai mechanism then we require $\Lplunge\gsim\Ltot>\Lz$. That is, according to the standard Kozai-Lidov formalism, for Kozai to drive a star to plunge, it must be true that
\BE
\Lz\lsim\Lplunge \;. \label{eqn:5:Lzplunge}
\EE
Stars with \Lz fulfilling this condition are said to lie inside the Kozai wedge \citep{Chen+09}.

%--------------------------------
\begin{figure}
\begin{center}
\includegraphics[width=\columnwidth]{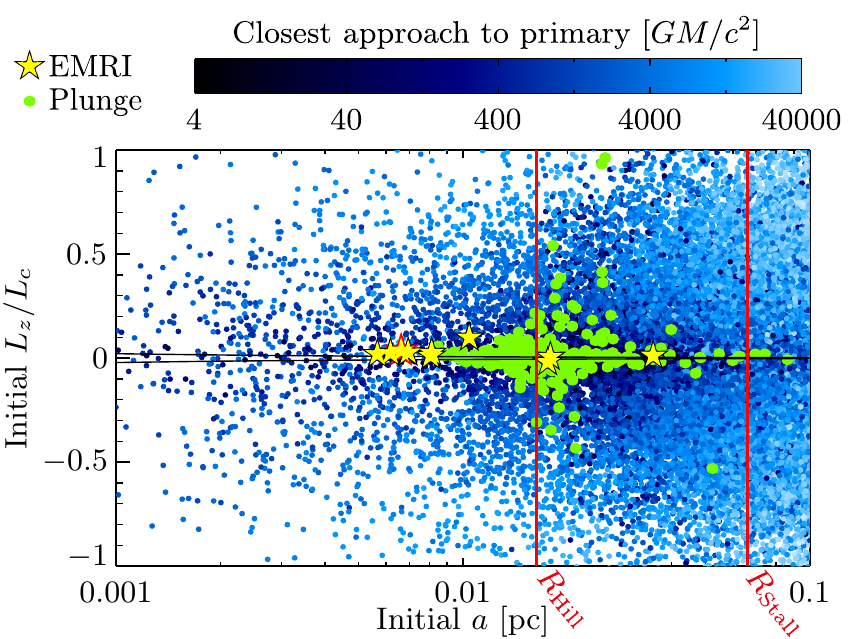}
\caption[We plot the initial $\Lz/\Lc$ as a function of the initial semi-major axis for the stars for the four major runs done]{\label{fig:5:kozaijc}
We plot the initial $\Lz/\Lc$ of our stars as a function of the initial semi-major axis, $a$. \Lc is the circular angular momentum: $\Lc^2 = G\mprim a$. In the Kozai formalism (\sect{sec:5:kozaiintro}) \Lz and $a$ are conserved over a star's evolution. The final stalling radius of the secondary is shown as a vertical red line. The simulation shown has $q=0.3$ and $\mstar=10\,\msun$. We encode the shortest distance to the primary SMBH a star reached over its evolution by colour: gold stars represent stars which have turned into EMRIs (\sect{sec:5:selectionandrejection}), green dots are stars which have plunged into the primary SMBH, and all other stars are coloured in shades of blue with darker blues being shorter distances (see legend, calculated using \eqn{eqn:5:trp}). Also plotted in solid black is the Kozai ``wedge''; lines of $\Lz=\pm4\, G\mprim /c$. Inside these lines the standard Kozai formalism (\sect{sec:5:kozaiintro}) predicts that it is possible, but not necessary, that a star reaches $\Ltot=4$ $G\mprim /c^2$. Thus, it is only inside these wedges that the standard Kozai formalism would predict plunges. Though this condition is not met for the majority of plunges/EMRIs, the clustering around small \Lz and at semi-major axes much smaller than the innermost position of the secondary indicate that the Kozai mechanism is important. That many EMRIs and plunges lie outside the Kozai wedge is primarily due to the oscillations on the orbital timescale of the secondary SMBH (see \sect{sec:5:fluctuationsinLzbin}), and is demonstrated in greater detail in \fig{fig:5:kozai}.
}
\end{center}
\end{figure}
%--------------------------------

The importance of this point is illustrated in \fig{fig:5:kozaijc} where we plot the outcomes of all stars in the simulation with $q=0.3$ and $\mstar=10\msun$ as a function of their initial $\Lz/\Lc$ and $a$, where \Lc is the circular angular momentum. Most plunges and EMRIs originate far from the secondary and come from a relatively narrow region of $\Lz/\Lc$, indicating that the Kozai mechanism is important to their evolution. 

However, the standard Kozai formalism does not explain the distribution of plunges, and EMRIs, in $\Lz/\Lc$. To illustrate this in \fig{fig:5:kozaijc} we plot the Kozai wedge (lines $\Lz=\pm4\, G\mprim /c$). The reason that most EMRIs and plunges are initially outside (but close to) the Kozai wedge is explained in \sect{sec:5:fluctuationsinLzbin}.

Note that \citet{Lithwick:Naoz:11} and \citet{Katz:11} have recently shown non-conservation of $L_z$ by the Kozai effect for the restricted three body case due to the octupole term in the perturbation expansion. In our simulations however, the secondary stalls with a low eccentricity ($e \sim 0.03$), and therefore the octupole term does not secularly perturb the orbit on the timescales simulated. They may however be important on longer timescales and with higher eccentricity perturbers.
%============================
\subsection{Kozai Effect --- Instantaneous Kozai Timescale}
\label{sec:5:kozaiadditionalpoints}
%============================

In addition to the Kozai timescale, \tKoz, we define a second `instantaneous Kozai timescale', \tkoz. \tKoz is the characteristic timescale describing the total period of of a Kozai oscillation. We define \tkoz to be the time that it takes for the orbital angular momentum of the star to change by its own magnitude. Thus, it allows one to understand the relevant timescale for change \emph{during} an oscillation. This is particularly useful for understanding when GR precession truncates an oscillation at high eccentricity.

The instantaneous Kozai-Lidov timescale is, to order unity, given by \citep{Chen+11}
\BE
\frac{1}{\tkoz} \sim  \frac{1}{\Ltot} \frac{d \Ltot}{d t} \,, \label{eqn:5:tkozinstantaneous}
\EE
where the specific angular momentum is given by 
\BE
\Ltot = \sqrt{G \mprim \astar (1-e^2)} \,,
\EE
and the torque due to the quadrupolar tidal force from the secondary is  
\BE
\left| \frac{d \Lvec}{d t} \right| = \left|\vec{F} \times \vec{r} \right| \sim \frac{q G\mprim \astar^2}{\asec^3} \,. \label{eqn:5:dLdt}
\EE
Together \eqns{eqn:5:tkozinstantaneous}{eqn:5:dLdt} give an instantaneous Kozai timescale of
\BA
\tkoz 
&\sim \frac{\sqrt{1-e^2}}{2 \pi q} \left(\frac{\asec}{\astar}\right)^3 \pstar \label{eqn:5:instantaneoustkozai} \\
&\sim \frac{\sqrt{1-e^2}}{2 \pi q} \frac{\psec^2}{\pstar} \ . 
\EA
Up to a constant and the factor $\sqrt{1-e^2}$ this is the standard Kozai timescale given by \eqn{eqn:5:Tkozai}. The non-constant factor $\sqrt{1-e^2}$ shows that the timescale for change in the angular momentum is shorter during periods of high eccentricity. This is because during periods of high eccentricity the orbit has the lowest angular momentum, requiring smaller torques to be significantly altered.

%============================
\subsection{Apsidal Precession}
\label{sec:5:apsidalprecession}
%============================

When any form of apsidal precession becomes comparable to \tkoz, then the magnitude of the Kozai oscillations is inhibited. This process is sometimes referred to in the literature as the Kozai mechanism being `de-tuned' \cite[e.g.][]{Thompson:10}. In our context there are two relevant forms of precession which affect the Kozai mechanism: that due to the stellar potential, and that due to GR precession.

The precession due to the non-Keplerian stellar potential results in an apsidal precession per stellar orbit of approximately \citep[e.g.][]{Merritt+11b}
\BE
\delta \omega_{\rm SP} \sim -2\pi \frac{\sqrt{1-e^2}}{1+\sqrt{1-e^2}} \frac{\Mstar(<\astar)}{\mprim} \,. \label{eqn:5:domegasp}
\EE
Hence, the timescale to precess though $\pi$ radians is
\BE
\tphisp \equiv \left| \frac{\pi}{\delta \omega_{\rm SP}} \right| \pstar \sim  {1\over2}\frac{1 + \sqrt{1-e^2}}{\sqrt{1-e^2}} \frac{\mprim}{\Mstar(<\astar)} \pstar \ .  \label{eqn:5:tphisp}
\EE
The mass ratio depends on the cusp model as discussed in \sect{sec:5:stellardistribution}.

On the other hand, GR precession has a different dependence. In the far field limit, the per orbit GR precession of a star is given by
\BE
\delta \omega_{\rm GR} = \frac{6 \pi G \mprim}{c^2 \astar (1-e^2)} = {3\pi\over1-e^2}{\rs\over\astar} \ . \label{eqn:5:domegagr}
\EE
Then, the timescale to precess through $\pi$ radians is given by
\BE
\tphigr=\left|\frac{\pi}{\delta \omega_{\rm GR}}\right| \pstar =  {1\over3}(1-e^2){\astar\over\rs}\pstar \ . \label{eqn:5:tphigr} 
\EE

It is important to note both that GR precession and SP precession are in opposite directions (\eqnsand{eqn:5:domegasp}{eqn:5:domegagr}) and that as the eccentricity of a star's orbit grows, $\tphisp$ increases while $\tphigr$ decreases (\eqnsand{eqn:5:tphisp}{eqn:5:tphigr}). 

We can see the individual effects in \fig{fig:5:revkozaiprecession} where, in the first column we plot the orbital parameters for the example star without either the effects of the stellar potential or GR, in the second column we plot the evolution including only the stellar potential, and in the third column we show the evolution of the full simulation with both the stellar potential and GR effects. One sees that in this case without the GR effects included the stellar potential largely damps the Kozai oscillations, while when they are included the star still reaches the angular momentum expected from the traditional Kozai formalism.

%============================
\subsection{Extreme Apsidal Precession}
\label{sec:5:extremeapsidalprecession}
%============================

When some non-Keplarian effect, other than Kozai, causes orbital precession on a timescale $\texterior \equiv \frac{2 \pi}{\dot{\omega}_{\rm ext}}$ which is much shorter than $\tKoz$, then the standard Kozai cycles are truncated on this new shorter timescale. Because $\texterior\ll\tKoz$, the argument of periapsis ceases to evolve according to \eqn{eqn:5:dwdt}, and instead follows an evolution dictated by this exterior effect (i.e.,  $\omega(t)\approx \int \dot{\omega}_{\rm ext}\,dt$). The $\sin2\omega$ term in the evolution of $e$ and $i$ (\eqnsand{eqn:5:dedt}{eqn:5:didt}) mean that both $e$ and $i$ will undergo two oscillations over a time $\texterior$, as expected for the quadrupole perturbation induced by the secondary. 

Moreover, given a star with eccentricity $e$ and angular momentum $L(e)$ this reduces the amplitude of the Kozai oscillations to roughly
\BE
\Delta L \sim L(e){\texterior\over\tkoz}\,.
\EE
That is, stars with low eccentricities will continue to have low eccentricity, but stars with high eccentricity retain their high eccentricity.

In the latter case, if a star reaches a high enough eccentricity that GR precession alone causes the orbit to precess significantly (e.g., by $\pi$ radians) before the angular momentum can change significantly (e.g., by its own magnitude) then the Kozai oscillations will be stalled at high eccentricity. When this occurs can be found by taking the ratio of $\tphigr$ to $\tkoz$,
\BE
{\tphigr\over\tkoz} \sim {2\sqrt{2}} q \left({\astar\over\asec}\right)^3{\astar\over\rs}\sqrt{1-e}\,,
\EE
and solving for $1-e$
\BE
1-e = {1\over 8}{1\over q^2}\left({\asec\over\astar}\right)^6\left({\rs\over\astar}\right)^2 \,.
\EE
This has been referred to as the Schwarzschild barrier in the context of resonant relaxation around single SMBHs by \citet{Merritt+11b}. In the case of our example star (shown in \figlistfive{fig:5:emriparameterspace}{fig:5:revkozaiprecession}{fig:5:kozaijc}{fig:5:kozai}{fig:5:oom}) this limit to the eccentricity occurs when $1-e\approx 3\times 10^{-4}$. In fact, this star does reach this eccentricity, subsequently ceases to oscillate, and forming an EMRI on a timescale too short to be clearly visible in \fig{fig:5:revkozaiprecession}.

Thus the example star highlights that the case of high eccentricity is of particular importance, since retaining high eccentricity accelerates the rate at which a star inspirals due to GW radiation. 

%============================
\subsection[Fluctuations in \Lz on the orbital timescale of the SMBH binary]{Fluctuations in \Lz on the orbital timescale of the SMBH binary\footnote{We note that an earlier version of this manuscript was shared with  \citet{Antognini:13} and \citet{Katz:12} who then nicely showed this effect can be important in hierarchical triples \citep{Antognini:13}, and mergers of white dwarfs in particular \citep{Katz:12}.}}
\label{sec:5:fluctuationsinLzbin}
%============================

In the standard Kozai mechanism, when averaged over the timescale of the SMBH binary, the component of angular momentum perpendicular to the binary's orbit, \Lz, is conserved. However, on shorter timescales this is not the case. This is because the symmetry about the $z$-axis is broken on shorter timescales. 

To illustrate this consider a short period over which the secondary SMBH moves negligibly: in this case the symmetry axis  of the quadrupolar tidal force on the star is directed towards the secondary ({\it not} in the $z$ direction). Therefore, it is this component of the angular momentum (which is perpendicular to $z$) that is conserved on very short timescales.

The resultant size of the fluctuations over the binary orbital period will be of order\footnote{A more precise but complex calculation of $\Delta \Lbin$ can be found in Appendix B of \citet{Ivanov+05}.}
\BE
\Delta \Lbin \sim \frac{d \Ltot}{dt} \frac{\psec}{4}
\label{eq:deltalb1}
\EE
where the factor of $4$ is to approximately take account of the dominant quadrupolar force, which gives rise to four reverses in sign per orbital period, \psec. Using the previously calculated torque (\eqn{eqn:5:dLdt}) we find
\BEA
\Delta \Lbin \sim \frac{q G\mprim \astar^2}{\asec^3} \frac{\psec}{4} &=& {\pi\over2}q\left({\astar\over\asec}\right)^{3/2}\sqrt{G\mprim\astar} \ , \label{eq:deltalb} \\
&\equiv& {\pi\over2}q\left({\astar\over\asec}\right)^{3/2} \Lc \,,
\EEA
so that
\BE
\Delta(1-e) = \frac{\Delta \Lbin}{\Lc} = {\pi\over 2}q{\pstar\over\psec}\,.
\EE
Here $\Lc=\sqrt{G\mprim \astar}$ is the maximum angular momentum with semi-major axis \astar, the circular angular momentum.

The fluctuations in $\Delta \Lbin$ become vitally important when $\Delta(1-e)\lsim1-e$. 
\BE
1-e\lsim{\pi\over2}q{\pstar\over\psec}={\pi\over2}q\left({\asec\over\astar}\right)^{3/2} \,.\label{eqn:5:delta1me}
\EE
This requirement is analogous to requiring that the variations in the periapsis distance due to the $\Delta \Lbin$ are comparable to the periapsis distance.

If the condition $\Delta(1-e)\lsim1-e$ is violated then the secular approximation breaks down. Therefore, whenever the secular Kozai equations are used, it is important to check that, at the highest eccentricities of interest, the size of the oscillations in eccentricity. 

%---------------------------------
\begin{figure}
\begin{center}
\includegraphics[width=\columnwidth]{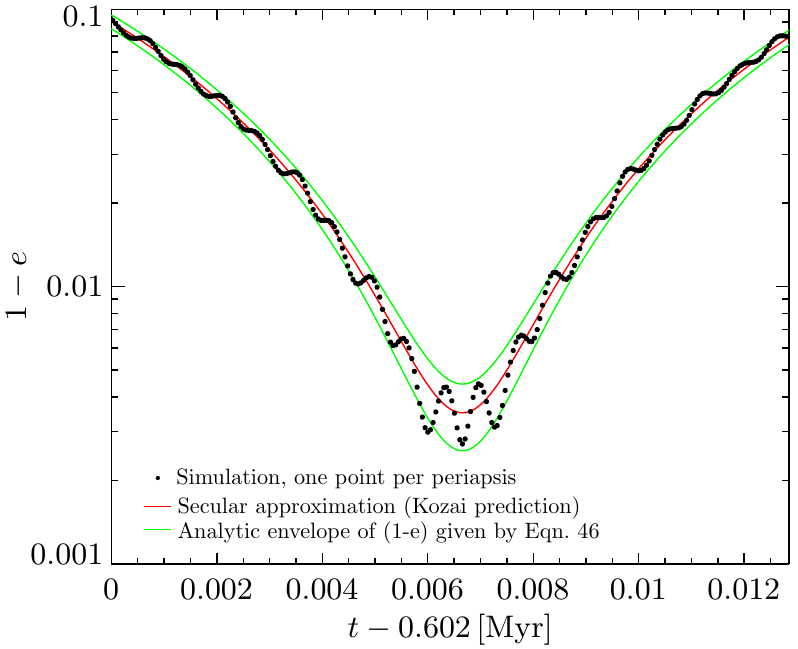}
\caption[We plot $1-e$ as a function of time across the peak of the oscillation in eccentricity.]{\label{fig:5:binaryoscillation} We plot $1-e$ as a function of time across the peak of the oscillation in eccentricity. In red is the approximate value predicted by the standard Kozai formalism (averaged over the secondary's orbit, \eqns{eqn:5:dadt}{eqn:5:dchidt}), while in green is the expected envelope of oscillations in $1-e$ over the secondary's orbital timescale given by \eqn{eqn:5:delta1me}. Each dot represents the value of $1-e$ calculated by our simulation at apoapsis during the first maximum of the eccentricity in the simulation shown in the first column of \fig{fig:5:revkozaiprecession} (in which both the stellar potential and the relativistic precession were turned off for clarity). The orbital modulations at the period of the secondary can be significant in these regions.}
\end{center}
\end{figure}
%---------------------------------

We demonstrate this effect in \fig{fig:5:binaryoscillation}. There we plot $1-e$ for a minimum of $\astar(1-e)$ in column 1 of \fig{fig:5:revkozaiprecession} as a function of time. In this case the star is being evolved without the effects of GR or precession due to the stellar potential so as to best illustrate the effect. In red is the approximate path predicted by the Kozai formalism, \eqns{eqn:5:dadt}{eqn:5:dchidt}, and in green is the predicted envelope given by \eqn{eqn:5:delta1me}. Each dot represents the calculated value of $1-e$ at periapsis during the integration.

The importance of these oscillations to the evolution of stars in our simulations is best demonstrated in \fig{fig:5:kozai}. This is  \fig{fig:5:kozaijc} with a non-normalized azimuthal angular momentum. In this plot the importance of region where $\Lplunge + \Delta \Lbin$ is clear and it approximately bounds most EMRIs and plunges. This is because stars with $\Lz < \Lplunge + \Delta \Lbin$ region are those that can reach \Lplunge via the Kozai effect, and therefore plunge into the primary SMBH or become EMRIs. 

%--------------------------------
\begin{figure}
\begin{center}
\roughdraft{\includegraphics[width=\columnwidth]{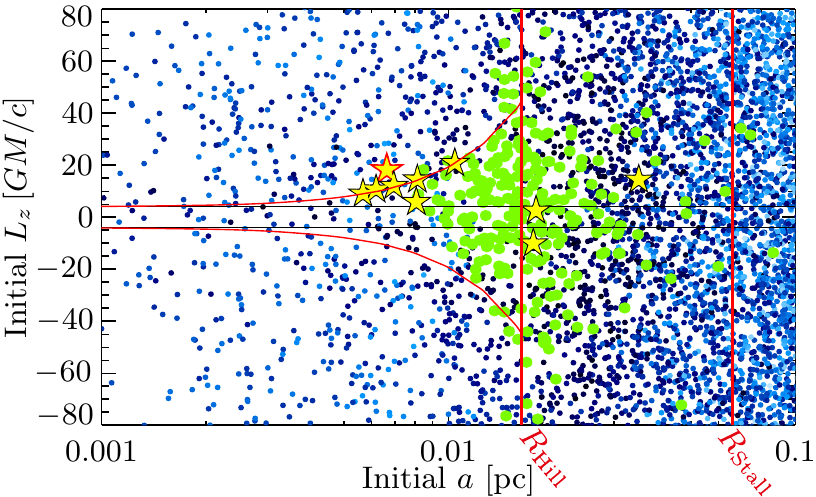}}{\includegraphics[width=\columnwidth]{./Figures/kozai.pdf}} 
\caption[We plot \Lz in units of $G\mprim /c$ as a function of the star's initial semi-major axis for the four major runs executed.]{\label{fig:5:kozai}
We plot the initial \Lz as a function of the stars' initial semi-major axes. The symbols and data sets are the same as in \fig{fig:5:kozaijc}. Here we also show in red $\Lplunge + \Delta \Lbin$ where $\Delta \Lbin$ are the range of possible oscillations of \Lz on the SMBH orbital timescale.  As is visible, these oscillations are important for most of the plunges and EMRIs. There is also a visible preference for driving stars with positive \Lz to become EMRIs or plunges, which is discussed briefly in the text.} 
\end{center}
\end{figure}
%--------------------------------

Interestingly, there is also a clear asymmetry about $\Lz=0$. This is a result of the a symmetry broken by the handedness of the secondary SMBH. When the orbit of the star is in the same sense as that of the secondary SMBH the star is more likely to reach higher eccentricities and ultimately become a plunge or EMRI. This is likely due to increasing the apparent period of the secondary SMBH during prograde GR precession. This increases the duration over which the torques from the binary are exerted and ultimately the total magnitude of $\Delta \Lbin$.

%============================
\subsection{Changes in angular momentum on the orbital timescale of the star}
\label{sec:5:fluctuationsinLzstar}
%============================

Between periapsis passages a star will undergo a change in angular momentum which is typically of size
\BE
\Delta \Lstar \sim \frac{d \Ltot}{dt} \pstar \sim \frac{q G\mprim \astar^2}{\asec^3} \pstar = \frac{G\mprim q}{\astar} \left(\frac{\astar}{\asec}\right)^{3}  \pstar \,,
\EE
which gives
\BE
\frac{\Delta \Lstar}{\Lc} = 2\pi q \left(\frac{\astar}{\asec}\right)^{3}=2\pi q\left({\pstar\over\psec}\right)^2 \,.
\EE
This is naturally of order $\pstar/\psec$ smaller than the oscillations on the SMBH binary timescale given by \eqn{eq:deltalb}. The importance in these oscillations is that while they remain small the star approaches \Lplunge more smoothly i.e. the discrete periapsis passages are closely spaced in angular momentum and periapsis distance. This is elucidated in the following subsection.

%============================
\subsection{Parameter Space}
\label{sec:5:parameterspace}
%============================

%--------------------------------
\begin{figure}
\centering
\includegraphics[width=\columnwidth]{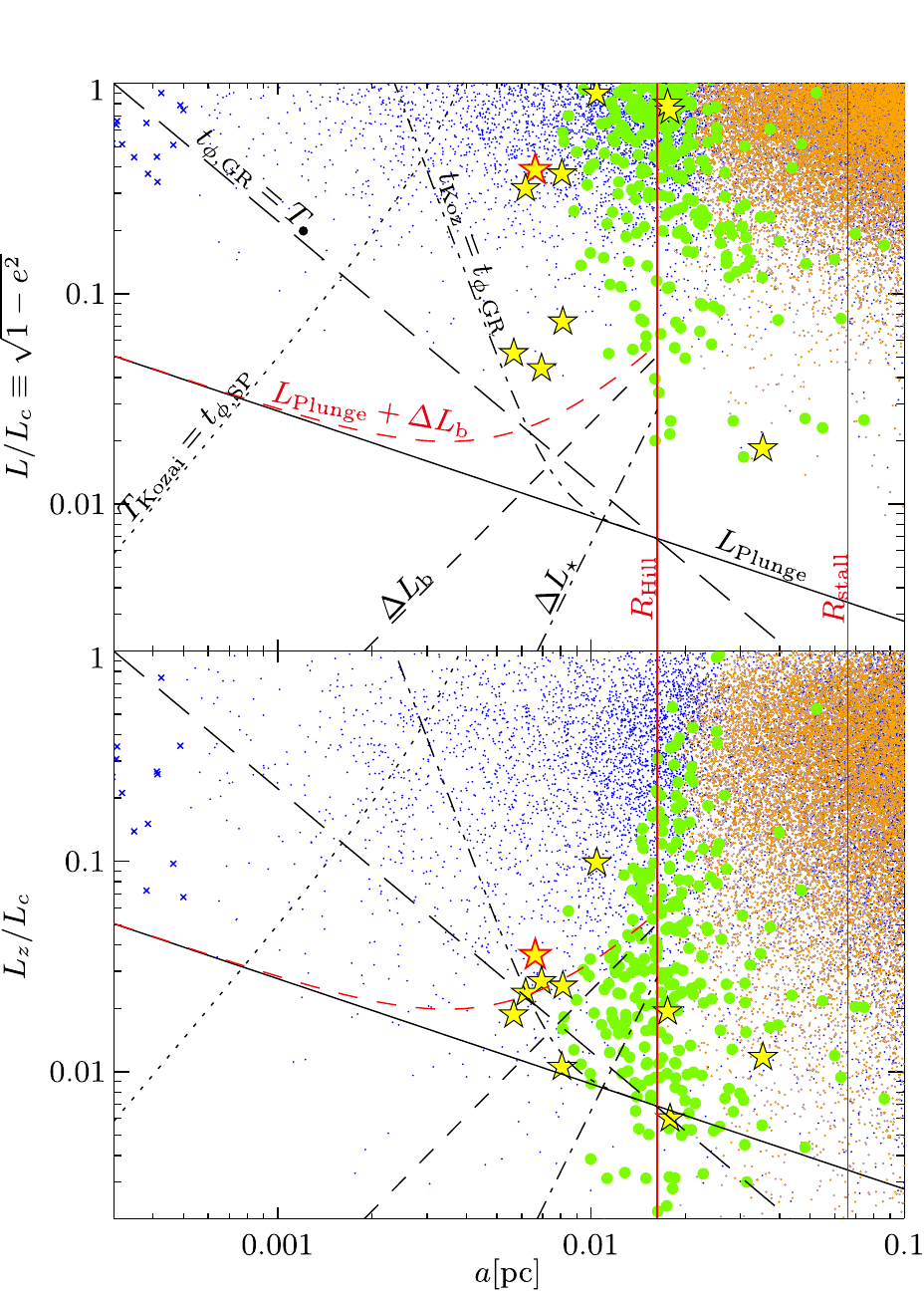}
\caption{A parameter space plot showing outcomes as function of normalised angular momentum and semi-major axis along with important delimiting lines as discussed in \sect{sec:5:oom}. Initial $\Ltot/\Lc$ and $\Lz/\Lc$ are plotted in the upper and lower panels respectively. Each star is represented by its outcome: an EMRI (gold star), a plunge (green dot), becoming unbound (orange dot), reaching our computational limit of $10^{10}$ steps (blue cross), or remaining in the cusp at the conclusion of our simulation (blue dot). The particular simulation shown is $q=0.3$ and $\mstar=10\msun$.} 
\label{fig:5:oom}
\end{figure}
%--------------------------------

To understand the effects of the various mechanisms discussed in this section, we plot the limits that they define in \fig{fig:5:oom}. The initial angular momentum, normalized by \Lc, is plotted against the initial semi-major axis of the stars in our simulation with $q=0.3$ and $\mstar=10\msun$. In the upper panel we plot the initial {\it total} angular momentum of the stars, while in the lower panel we plot the initial $z$ component of the angular momentum. Stars that result in EMRIs are demarcated by gold stars, stars that plunge by green dots, stars that become unbound by orange dots, and all other stars by blue dots.

The secondary stalls at the stalling radius labeled $\rstall$. At this point the primary has a resultant Hill radius labeled $R_{\rm Hill}$. We define the Hill radius around the primary as the radius inside of which the Jacobi constant cannot be small enough to cross the L$_2$ Lagrange point beyond the secondary and exit the system \citep{Murray:Dermott:00}. The stars that are unbound and ejected from the system are orange and restricted to semi-major axes greater than the Hill radius. 

Lines are shown where the timescales estimated above are equal: $\tkoz=\tphisp$ (dotted black), $\tkoz=\tphigr$ (solid black diagonal), $\psec=\tphigr$ (long dashed black). Also plotted are lines demonstrating the size of changes in angular momentum during the binary's orbit $\Delta \Lbin$ (short dashed black), and over an orbit of the stars $\Delta \Lstar$ (dash dot black). For orbits close with \Ltot close to \Lplunge then the far field approximation of \eqn{eqn:5:domegagr} is inaccurate. This is the reason for the curvature of the $\tkoz=\tphigr $ line, which is calculated using the geodesic equation.

Near or outside the Hill radius stars evolve by strong interactions with the secondary SMBH and the Kozai effect is not relevant. As is visible from the figure it is less probable that these ``chaotic orbits'' turn into plunges or EMRIs than stars farther in which are affected by the Kozai mechanism.

Well within the Hill radius the Kozai mechanism is most accurate. By the standard Kozai formalism no star with an initial \Lz much greater than \Lplunge should plunge (or form an EMRI), since \Lz is conserved. However, due to oscillations in \Ltot, ($\Delta \Lbin$) which occur on the binary orbital timescale and are not accounted for in the Kozai formalism stars with higher \Lz can still become plunges or EMRIs if $\Lz \lsim \Lplunge + \Delta \Lbin$.

The plunges and EMRIs lie to the right of the Schwarzschild barrier ($\tkoz = \tphigr$) since to the left of this line GR precession dominates the evolution and it is not possible to reach low angular momenta before the orbit precesses and the tidal torque is reversed. 

Moreover, in the region with $a$ small but to the right of $\tkoz = \tphigr$, the changes of angular momentum over each orbit ($\Delta \Lstar$) are small and therefore stars gradually approach \Lplunge. As a result these stars undergo close periapsis passages and lose their energy to gravitational radiation instead of directly plunging. To the right of this region, with larger $a$, the change in angular momentum on each orbit becomes larger, and stars are more likely to plunge directly into the primary than become EMRIs.

%============================
\subsection{Synopsis}
\label{sec:5:synopsis}
%============================

There are essentially three different ways of producing EMRIs in the context of widely-separated binary SMBHs: 1) single or multiple strong interactions with the secondary SMBH which cause the compact object to fortuitously pass close to the primary; 2) Kozai oscillations mixed with oscillations on the timescale of the secondary which drive the star to high eccentricity and has some significant probability of plunging but instead has a close passage; and 3) stars which are also driven to high eccentricity from Kozai oscillations mixed with oscillations on the timescale of the secondary, but approach the Schwarzschild barrier before, but in the near vicinity of \Lplunge. In our simulations the first of these methods is sub-dominant, and so we focus on the latter two.

In both cases the initial process is the same and so we shall describe them together. The star must begin with a low \Lz such that $\Lz\le\Lplunge+\Delta\Lbin$. This alone is necessary but not sufficient to ensure that a star will reach \Lplunge. Secondly, the star should have a semi-major axis which is at most a factor of a few smaller than \rstall. This avoids the strong scatterings by the secondary that are most likely to eject the star from the system, and instead results in Kozai-Lidov oscillations. However, the semi-major axis of the star can not be so low that the Kozai period becomes too long and GR precession dominates the evolution ($\tkoz > \tphigr$) before reaching low angular momenta. In this case the orbit will precess and the tidal torque that drives the Kozai-Lidov oscillations will be reversed well before \Lplunge. GW radiation therefore cannot act efficiently to drive the star to be an EMRI.

In our simulations we find a `sweet spot' where the semi-major axis is just larger than the Schwarzschild barrier. In this region the changes in angular momentum between periapsis passages are small ($\Delta \Lstar \ll \Lplunge$). The star therefore has close periapsis passages in which it can radiate orbital energy in GWs, forming an EMRI before plunging beyond the Schwarzschild radius. At larger semi-major axes the change in periapsis distance per orbit is larger, and stars are more likely to plunge directly into the primary SMBH.

%=============================================
\section{Discussion} 
\label{sec:5:discussion}
%=============================================

%----------------------------------------------------------------------------
\subsection{EMRI Rates}
\label{sec:5:emriratesdiscussion}
%----------------------------------------------------------------------------

Under the previous design of LISA EMRIs consisting of a $10^6\msun+10\msun$ black hole system would be detectable to a redshift of $z \sim 1$ \cite[dimensionless spin $a/M=0.9$, averaged over orientations,][]{AmaroSeoane+07} giving a co-moving detection volume of $160\Gpc^3$. The rate of EMRIs from SMBH binaries estimated here of $\mathcal{R}_{\rm EMRI}=8\times10^{-4}\yr^{-1}\Gpc^{-3}$ (table \ref{tab:5:finalrates}), gives a detection rate of $\sim 0.12\yr^{-1}$, and therefore $\sim 0.6$ over a five year mission.

This is particularly interesting since EMRI waveforms contain information about the presence of both gas \citep{Narayan:00,Yunes+11} and the secondary SMBH \citep{Yunes+10}. The importance of gas is significant, since due to the same mechanisms discussed here, a large number of tidal disruptions are also expected \citep{Chen+08,Wegg:Bode:10,Chen+11}. Thus, any EMRI observed with the signal of a secondary SMBH in its waveform formed by the mechanisms discussed here are likely to be in the presence of gas.

It is important to note that our rates are proportional to the number density of compact objects at about $\rstall/10\sim 0.01\,{\rm pc}$. Note that this is in contrast to the standard picture of isolated SMBH EMRI formation where the rate scales with the product of the number density of compact objects and that of stars \citep[e.g.][]{Hopman:09}. Therefore, in the fully mass segregated case, the isolated SMBH EMRI formation rate scales with the square of the number density of compact objects. We discuss some of our assumptions which affect the number density of compact objects in \sect{sec:5:assumptions}.

There is another key consequence of the EMRI formation mechanism presented here, which is outside the scope of this work: In the standard picture of EMRI formation stars must scatter to a state with low overall \Ltot, while in the Kozai picture stars need only a low \Lz (i.e. the standard loss cone is instead the entire Kozai wedge, see \figsand{fig:5:kozaijc}{fig:5:kozai}). Thus, for the duration of the secondary's time at the stalling radius, interactions  \citep[such as star-star scattering, or non-conservation of \Lz due to the octopole term][]{Lithwick:Naoz:11} need only drive stars to low \Lz (or more accurately low $\Lz+\Delta \Lbin$, see \sect{sec:5:parameterspace}) for them to be able to form EMRIs under the Kozai mechanism. 
This situation is similar to the predicted increased rates of tidal disruption in axisymmetric nuclei. In this case the rates are increased by a factor of a few from spherical nuclei \citep[]{Magorrian:99,Vasiliev:13}. We therefore conservatively expect an increase of at least this factor in the EMRI rate while the secondary SMBH is stalled. However there is at least one reason to suspect rate increase would be higher than this: The oscillations on the timescale of the SMBH binary (\sect{sec:5:fluctuationsinLzbin}) do not occur in axisymmetric nuclei, and these expand the size of the loss cone. The situation warrants further investigation but is beyond the scope of this work.

%----------------------------------------------------------------------------
\subsection{Assumptions}
\label{sec:5:assumptions}
%----------------------------------------------------------------------------

{\it Schwarzschild Black Holes:}  Throughout this work, in common with the vast majority of studies on the formation of EMRIs, we neglect the spin of the SMBH. This is a short coming that is only beginning to be overcome \citep{AmaroSeoane:12}. We  make this choice for simplicity, since there is increasing observational evidence that at least some SMBHs have significant spin \citep{Brenneman:Reynolds:06}. 
Apart from very close passages, even in the presence of spin, the precession will be dominated by the Schwarzschild terms \citep{Merritt+09}. Instead the largest effect on this work would be Lense-Thirring precession of the star's orbital plane since, if the BH spin is not aligned with the orbital plane, this would result in non-conservation of \Lz. We speculate this could therefore increase the rate of EMRIs and plunges since more stars can potentially undergo Kozai oscillations which result in close BH encounters. 
To lowest order in $v/c$ the angular momentum of a test particle in the Kerr metric precesses due to Lense-Thirring precession at a rate \citep[][we neglect the quadrupolar term in this order of magnitude estimate]{Merritt:Book}
\BE
\frac{d \bmath{L}}{dt} = \frac{4\pi}{\pstar} \left( \frac{GM}{Lc} \right)^3 (\bmath{\chi} \times \bmath{L})
\EE
where $\bmath{\chi}$ is the dimensionless spin vector of the black hole. Over $n$ orbits we therefore expect a change in $\Lz$ of
\BE
\Delta \Lz \approx 4\pi n \left( \frac{GM}{Lc} \right)^3 (\bmath{\chi} \times \bmath{L})_z
\EE
The majority of the EMRIs have initial semi-major axis $a\approx 0.007\pc$ and have $n\approx 20,000$ orbits over the entire length of our simulations. For a favourably oriented spinning hole then, over our simulation, stars at this semi-major axis change their \Lz by
\BE
\frac{\Delta \Lz}{GM/c} \approx \frac{1.7 \chi}{1-e^2} ~.
\EE
Therefore even over the $\approx 1\Myr$ length of our simulations a significant fraction of stars at this radius could have their \Lz secularly changed by of order the size of the loss cone by a spinning black hole. These perturbations would continue for the entire length of time that the binary is stalled.

{\it Cusp Profile:} One of the major factors in determining the rates is the stellar distribution. In our simulations we use an $\eta$-model \citep{Tremaine+94} of a spherical stellar cusp with a central SMBH to establish the stellar distribution (see \sect{sec:5:stellardistribution}). This is a self-consistent family of models of a stable isotropic stellar cusp. In our simulations, we have chosen $\eta=1.75$, the value appropriate for a relaxed stellar cusp. However, there is a complication: the  galaxy where we can best resolve the inner parsec is our own Milky Way, and as yet there is no consensus on the existence of a  cusp \citep{Buchholz+09,Bartko+10,Do+09,YusefZadeh+12}. An alternative interpretation to a lack of a steep visible cusp is that a density cusp is present in the Galactic center, but is `dark' as a result of mass segregation causing the density to be dominated by COs \citep{Freitag+06,Preto:AmaroSeoane:10}. Our rates scale roughly linearly with the number density of stellar mass black holes at $R_{\rm stall}/10\sim 0.01\,{\rm pc}$, allowing them to easily be rescaled to other cusp profiles (and other CO number densities).

{\it Stellar Interactions:}  We have not considered relaxation processes such as those due to star-star scattering or star-bulk scattering. 

The timescale for relaxation via star-star scattering is approximately $1~{\rm Gyr}$ at $r_c$ \citep{AmaroSeoane:Preto:12}, much longer than the duration of our simulations, and is not a strong function of $r$ in the cusp \citep{Alexander:05}. However, care must be taken: This approximation is not as accurate as might be assumed because the timescale to change angular momentum by of order itself will be reduced for high eccentricity orbits by a factor $\sim (1-e^2)$~\citep{Hopman:Alexander:05}. Therefore, for the highest eccentricity stars in our simulation relaxation could be beginning to become non-negligible. 

In the case of star-bulk relaxation such as that due to resonant relaxation or asymmetric bulges the timescales can be much shorter. For instance, consider resonant relaxation. In this case the comparable timescale to \tkoz is given by \citep{Merritt+11}
\BE
\trr=\sqrt{N_\star(<\astar)}{\mprim\over M_\star(<\astar)}{P_\star\over 2\pi}\sqrt{1-e^2} \,.
\EE
Here $N_\star(<\astar)$ is the number of stars inside the semi-major axis of the test star. Equating \trr to \tkoz we can solve for the semi-major axis where the two effects are comparable:
\BE
\arr = 0.004 \left({\rstall\over0.07\pc}\right)^{24/19}\left({q\over0.3}\right)^{-8/19}\left({\mprim/\mstar\over10^5}\right)^{-4/19} \pc \;.
\EE
Inside of $\arr$ resonant relaxation would be the dominant form of precession while further out the Kozai torques would dominate the evolution. While $\arr$ is about a factor of $2$ smaller than where our innermost EMRIs are sourced it could have an impact on our results. However, this is beyond the scope of this work.

%--------------------------------
\begin{figure}
\begin{center}
\includegraphics{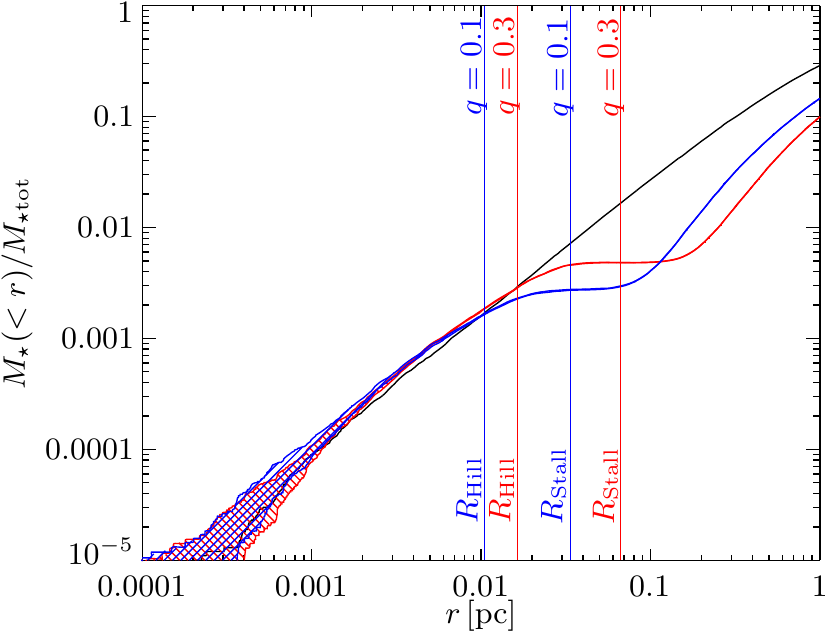}
\caption[We plot the stellar mass interior to a given radius.]{\label{fig:5:mintr}
We plot the stellar mass interior to a given radius normalized to the total stellar mass, $\Mstar(<r) / M_{\star \rm tot} $. In black we plot the initial cumulative stellar mass (c.f. \eqn{eqn:5:massinterior}), while in blue we plot the cumulative stellar mass at the end of the simulation for the $q=0.1$, $\mstar=10\,\msun$ run, and in red we plot the cumulative stellar mass at the end of the simulation for the $q=0.3$, $\mstar=10\,\msun$ run. Note that the initial conditions for all runs are the same. The only difference between the outcomes of the runs with the same $q$ but different stellar masses is due to those stars which go within $100\, G\mprim /c^2$, only a small fraction of all stars. Thus, the cumulative stellar mass for the runs with $\mstar=1\,\msun$ are virtually the same as the $10\,\msun$ counterparts with the same $q$. The hatched regions are stars that had not completed the full simulation within the preset limit of $10^{10}$ steps. For reference we also plot the stalling radii of the secondary as vertical lines.
}
\end{center}
\end{figure}
%--------------------------------

{\it SMBH merger rate:}
While in recent years there has been significant progress in understanding the merger rates of SMBHs, there still remains a great deal of uncertainty. Indeed, this is one of the unknowns that LISA or a LISA-like experiment would estimate. Here we have made the crude approximation that every $10^6\msun$ SMBH will undergo one major merger per Hubble time with a constant probability over time. While it is certain that neither of these assumptions is quite right, we expect the uncertainties here to be minimal when compared to those relating to the stellar cusp, or, potentially, the lack thereof. For this reason we have chosen simplicity over false precision.

{\it Invariant stellar potential:} 
One inconsistency of our methodology is the assumption that the stellar potential does not evolve with time, though the stars' orbits do. To demonstrate the possible effect of such an assumption we plot the mass interior to a given radius as a function of radius in \fig{fig:5:mintr}. There the solid black line is the initial distribution given by \eqn{eqn:5:massinterior}, the red solid line is the curve for the $q=0.3$ and $\mstar=10\,\msun$ simulation, and the blue solid curve is for the $q=0.1$ and $\mstar=10\,\msun$ simulation. The mass of the star has little effect on these curves. The filled region represents stars that required more than $10^{10}$ steps to complete the simulation and were therefore terminated. From \figsand{fig:5:kozaijc}{fig:5:kozai} most EMRIs originate from  $\approx 10^{-2}\,\pc$ and at this position the mass interior has not changed significantly. Thus, this assumption would not likely have a significant impact on our findings.

%=============================================
\section{Conclusion}
\label{sec:5:conclusion}
%=============================================

We have considered the possibility of extreme mass-ratio inspirals (EMRIs) that form as a result of a secondary supermassive black hole (SMBH) inspiraling towards a primary SMBH. Using a symplectic integrator to follow the paths of $10^6$ non-interacting stars around a primary $10^6\msun$ SMBH with various values for both the stellar mass and an inspiraling secondary SMBH, we have reached several conclusions:

\BEn
\item EMRIs can be formed by binary SMBH systems with numbers which could detectable by future space-based gravitational wave missions.
\item Frequently overlooked, oscillations on the timescale of the secondary SMBH, and which deviate from the traditional Kozai-Lidov mechanism, are fundamental to the formation of almost all EMRIs and plunges.
\item The region of parameter space where precession due to GR, \tphigr, and the precession due to the Kozai-Lidov mechanism, \tkoz, are comparable (the Schwarzschild barrier: $\tphigr=\tkoz$) causes important dynamical effects.
\item When a star has a semi-major axis such that $\tkoz<\tphigr$, neither EMRIs nor plunges are possible since the precession rate due to GR will increase before the angular momentum of the star falls sufficiently to plunge or become an EMRI.
\item When a star has a semi-major axis such that  $\tkoz>\tphigr$, EMRIs are formed most often when the star has a smaller semi-major axis, since it will gradually approach the plunge angular momentum (small $\Delta \Lstar$). At higher semi-major axes plunges are more likely.
\EEn

It is also interesting that the formation of EMRIs by the channel described here is verifiable: While we have shown that it is possible for EMRIs to be formed in SMBH binaries, it has also been shown that, if such an EMRI were to form and be detected, it is possible that the waveform provides information about the mass of the secondary SMBH and the binary separation \citep{Yunes+10}. This is of particular interest since, by inferring the existence of a SMBH binary at wide separation, it would extend the range of SMBH binary separations to which low frequency gravitational wave detectors are sensitive.

%-------------------------------------ACKNOWLEDGMENTS----------------------------------

%--------------------------------------------------------------------------------------
\section{Acknowledgments }
We gratefully acknowledge a significant amount of guidance from Sterl Phinney, many insights regarding resonant relaxation from David Merritt, and useful conversations with Sotiris Chatzopoulos, Chris Hirata, Smadar Naoz and Michele Vallisneri.

Support for this work was provided by NASA BEFS grant \supportfrom{NNX-07AH06G}.
%--------------------------------------------------------------------------------------

\bibliographystyle{mn2e}
%\bibliography{mn-jour,bibliography}

\label{lastpage}
\end{document}